\documentclass[12pt,preprint]{aastex}

\usepackage{graphicx}
\usepackage{enumerate}
\usepackage{amsmath}
\usepackage{comment}

\def\be{\begin{equation}}
\def\ee{\end{equation}}
\def\bea{\begin{eqnarray}}
\def\eea{\end{eqnarray}}
\def\bn{\begin{enumerate}}
\def\en{\end{enumerate}}

\def\sec\ond{{\rm s}}

\def\Mpc{{\rm Mpc}}
\def\hMpc{~\,h^{-1}\Mpc}

\def\ie{{\frenchspacing\it i.e.}}
\def\eg{{\frenchspacing\it e.g.}}

\def\rms{{\frenchspacing r.m.s.}}

\shorttitle{Systematic Effects}
\shortauthors{Kim et al.}
\begin{document}
\title{Systematic Effects on the Genus Topology of Large Scale Structure of the Universe}
\author{Young-Rae Kim\altaffilmark{1}, Yun-Young Choi\altaffilmark{2},  
Sungsoo S. Kim\altaffilmark{3}, Kap-Sung Kim\altaffilmark{2},
Jeong-Eun Lee\altaffilmark{3}, Jihye Shin\altaffilmark{3}, and
Minbae Kim\altaffilmark{3}}
\affil{
$^1$School of Physics, Korea Institute for Advanced Study, Heogiro 85, Seoul 130-722, Korea\\
$^2$Department of Astronomy and Space Science, Kyung Hee University, Gyeonggi 446-701, Korea\\
$^3$School of Space Research, Kyung Hee University, Gyeonggi 446-701, Korea}
\begin{abstract}
Large-scale structure of the universe is a useful cosmological probe
of the primordial non-Gaussianity and the expansion history of the universe
because its topology does not change with time in the linear regime
in the standard paradigm of structure formation.
However, 
when the topology of iso-density contour surfaces is measured 
from an observational data, many systematic effects are introduced 
due to the finite size of pixels used to define the density field, 
non-linear gravitational evolution, 
redshift-space distortion, shot noise (discrete sampling), 
and bias in the distribution of the density field tracers.
We study the various systematic effects on the genus curve to 
a great accuracy by using the Horizon Run 2 simulation of a $\Lambda$CDM cosmology.
We numerically measure the genus curve from the gravitationally evolved 
matter and dark matter halo density fields. It is found that 
all the non-Gaussian deviations due to the systematic effects can be modeled  
by using a few low-order Hermite polynomials from $H_0$ to $H_4$.
We compare our results with the analytic theories whenever possible, 
and find many new terms in the Hermite series that are making significant 
contributions to the non-Gaussian deviations.
In particular, it is found that the amplitude drop of the genus curve
due to the non-linear gravitational evolution can be accurately
modeled by two terms $H_0$ and $H_2$ with coefficients both proportional 
to $\sigma_0^2$, the mean-square density fluctuation.
\end{abstract}

\section{Introduction}
The topology of large-scale structures (LSS) of the universe has long been 
used as a probe of non-Gaussianity of the primordial density fluctuations 
(Gott et al. 1986; Matsubara 1994; Park et al. 2005; Hikage et al. 2006;
Hikaget et al. 2008). 
It has also been recently suggested to use 
the LSS topology as a cosmological invariant that can be used to reconstruct 
the expansion history of the universe (Park \& Kim 2010; Zunckel et al. 2011;
Wang et al. 2012; Blake et al. 2013; Speare et al. 2013).
The cosmological genus statistic has been the most popular measure 
of the LSS topology and used to constrain the primordial non-Gaussianity 
(Choi et al. 2013), galaxy formation mechanism, and cosmology 
(Park et al. 2005; Choi et al. 2010; Way et al. 2011 among many others).

The genus as applied in cosmology quantifies the connectivity of iso-density or 
iso-temperature contour surfaces of smooth matter density or cosmic microwave 
temperature field, and is equal to $-1/2$ times the Euler characteristic of the surface 
or equivalently a linear combination of the Betti numbers of the excursion sets, 
topological invariants of figures that can be used to distinguish topological spaces 
(see Park et al. 2013 for the relations among the genus, Euler characteristic, and the
Betti numbers in two- and three-dimensions).
The genus is also one of the parameters characterizing the geometry of figures called 
the Minkowski Functionals 
(Mecke et al. 1994; Pratten \& Munshi 2012; Hikage et al. 2006; Hikage et al. 2008). 

In the analysis of the topology of the observed LSS, as in many other cases, the genus
directly measured from the observational data contains large systematic effects.
It is thus critically important to accurately estimate the systematic effects
and statistical uncertainties in order to draw any sensible conclusions.
An example of extreme care of various systematic effects on the genus is the work by
Choi et al. (2010) and Choi et al. (2013) where, 
in addition to sophisticated correction for the radial and
angular selection effects in the observational data, a number of mock surveys performed
within a large cosmological simulation mimicking the actual observation
in detail are used to
simulate the systematic effects produced by non-linear gravitational evolution, 
redshift-space distortion, shot noise in the smoothed density field, bias in the
distribution of galaxies with respect to the underlying matter density field, and 
the finite size of the pixels used to build the density field.
Without taking into account of these effects one would make completely wrong
conclusions on the interpretation of the deviation of the measured genus curve from
the Gaussian predictions.  
The primordial non-Gaussianity can be studied only through statistical 
comparison of observations with the accurately modeled mock survey samples 
in a simulated universe with well-defined initial conditions.
Therefore, mock survey samples drawn from large-volume cosmological simulations
are an essential element of topology analysis of LSS that enables one to model
all the systematics and to correctly estimate the statistical uncertainties in
the derived cosmological parameters.

There have been a number of studies that tried to estimate some of these systematic
effects individually. The finite pixel size effects have been analytically studied for 
the 3-dimensional genus by Hamilton et al. (1986) and the 2-dimensional genus
by Melott et al. (1989). 
Melott et al. (1988) and Park \& Gott (1991) numerically studied the combined 
effects of the non-linear gravitational evolution and galaxy biasing.
An analytic formula for the effects of non-linear gravitational evolution has been
found by Matsubara (1994, 2003) using the second-order perturbation theory, 
which was confirmed by Matsubara \& Suto (1996) in the weakly nonlinear regime. 
The redshift-space distortion effects in the linear regime were found 
by Matsubara (1996).
James (2012) discussed the systematic effects inducing non-Gaussianity in the genus 
statistic produced by galaxy bias, nonlinear gravitational evolution and primordial
non-Gaussianity and presented a technique to decomposing them into an orthogonal
polynomial sequence. The effect of non-periodic sample boundaries in conjunction 
with smoothing method, which is not dealt with in this paper, was studied 
by Melott \& Dominik (1993).

In practice, the systematic effects in the genus measured from observations or mock
surveys are all combined and reveal themselves as a composite deviation from the
Gaussian curve. However, it is theoretically interesting to know the sign and
magnitude of individual effects. With such understanding one can make
a better control of individual effects and it would be possible to reduce them.
In this paper we will measure the individual effects of these systematics using
the matter density field and dark matter halo distribution obtained 
from a cosmological $N$-body simulation, the Horizon Run 2 (HR2), 
one of the largest cosmological simulations in terms of the number of 
particles evolved and the physical volume simulated (Kim et al. 2011).

The non-Gaussian deviations of the genus curve we will present 
are correct only for the particular cosmological model we adopt 
(a flat $\Lambda$CDM cosmology with the
5 year WMAP cosmological parameters) and the galaxy-subhalo
correspondence model of galaxy assignment we adopt in this paper. However, it is
believed that these models are reasonable approximations (for massive objects
in particular) and any small changes in cosmology and galaxy assignment scheme
would not change our conclusions on the nature of the non-Gaussian
genus deviations due to systematic effects.

The paper is structured as follows.
In Section 2 we briefly introduce the genus statistic.
In Section 3 we explain the numerical simulation used in our study.
Section 4 discusses the systematic effects and our modeling,
and the summary and conclusion are given in Section 5.

\section{The Genus Statistic}

The genus is a measure of the topology of iso-density
contour surfaces in a smoothed galaxy density field.
In cosmology the genus is usually defined as (Gott et al. 1986)
\begin{equation}
G={\rm number\; of\; holes}-{\rm number\; of\; isolated\; regions},
\end{equation}
where the ``number of isolated regions'' is the number of disconnected
pieces into which the boundary surface is divided and
the ``number of holes'' is the maximum number of cuts that can be
made to the regions without creating a new disconnected region.

In the case of Gaussian fields, the genus per unit volume
in the iso-density contour surface at a given threshold level $\nu$, $g(\nu) = G(\nu)/V$,
is known as (\eg Doroshkevich 1970; Adler 1981; Hamilton et al. 1986) 
\begin{equation}
\label{eq:GRgenus}
g(\nu) = A(1 - \nu^2)e^{-\nu^2/2}.
\end{equation}
The amplitude $A$ is given by
\begin{equation}
A = -{1\over{8\pi^2}}  \left( {{\langle k^2 \rangle }/3}\right)^{3/2},
\end{equation}
where $\langle k^2\rangle=\sigma^2_1/\sigma^2_0$. 
The spectral moments of the fields, $\sigma_0$ and $\sigma_1$
are computed from $\sigma_0=\sqrt{\left<\delta^2\right>},$ and 
$\sigma_1=\sqrt{\left<|\nabla \delta|^2\right>}$. 
$\delta=(\rho-{\bar \rho})/{\bar \rho}$ is the over-density field.

To separate the contribution of the one-point density distribution
to the genus the volume-fraction threshold level has often been used
in topology analyses (Weinberg et al. 1987).
This parameter defines the iso-density contour surfaces
in such a way that the volume fraction $f$ in the high density region
is equal to that of a Gaussian random field at the same
threshold level, where
\begin{equation}
f = {1\over\sqrt{2\pi}}\int_\nu^\infty e^{-x^2/2} \,dx.
\end{equation}
This transformation of the density field, called Gaussianization, in general reduces
the non-Gaussian features in the genus curve, but is useful when
the non-Gaussianity of the primordial density field is to be studied 
(see Neyrinck et al. 2011 for matter density field, and Jee et al. 2012 for
halo density field).
We will use this parametrization in this paper for easier comparison
with previous works, and
the parameter $\nu$ means the volume-fraction threshold.

\section{Simulation}

We use matter particle and dark halo data of an $N$-body simulation called Horizon
Run 2 (HR2) that gravitationally evolved $6000^3$ particles 
in a cubic comoving volume of $(7200\hMpc)^3$ (Kim et al. 2011). 
The simulation is based on the standard 
concordance model with parameters corresponding
to the WMAP 5-year cosmology (Komatsu et al. 2009) of
$\Omega_{\Lambda}=0.74$, $\Omega_{m}=0.26$, $\Omega_{b}=0.044$, $n_{s}=0.96$,
$h=0.72$, and $\sigma_{8}=0.794$. The particle mass is about $1.25\times10^{11}\hMpc$
and the mean particle separation is 1.2$\hMpc$.
The initial redshift of the simulation is $z_{\rm i}=32$. 
The initial particle distribution are generated by using the Zel'dovich 
approximation which adopts the first-order Lagrangian perturbation theory. 
Even though the small-scale ($k\geq 0.2\hMpc$) power
shows a subtle underestimation compared with the second-order scheme 
(L'Huillier et al. 2014), at low redshifts the Zel'dovich approximation 
does not show any noticeable difference for the smoothing scales 
that we are interested in here 
(Gaussian smoothing lengths from $R_G = 15$ to 34$\hMpc$)

We use the snapshot data of HR2 at $z=0$ in our study. The physical size of HR2 is large enough to
correctly model the growth of LSS through coupling of density fluctuations on various
scales and to allow us to measure all the systematics to great accuracies with negligible
statistical uncertainties.
To study the galaxy biasing effects on the genus, we adopt the subhalo-galaxy correspondence
model (Kim et al. 2008) and assign a galaxy at the location of each dark halo.
The dark matter halos are the physically self-bound (PSB) halos identified in HR2
using the method described in Kim \& Park (2006).
They are gravitationally self-bound, and tidally stable
halos hosting a density maximum.
Since a roughly volume-limited subsample of galaxies that can be extracted from the
on-going SDSS-III BOSS survey for topology analyses has the mean galaxy separation 
of about $15\hMpc$ (Choi et al. in preparation), we set the minimum halo mass of 
$1.61\times 10^{13}~h^{-1}{\rm M_\odot}$ (129 particles)
so that the average PSB halo separation in HR2 becomes also 15$\hMpc$
This subhalo-galaxy correspondence model has been proven to work well in terms of 
one-point function and its local density dependence (Kim et al. 2008), two-point function 
(Kim et al. 2009), and also topology in particular (Gott et al. 2009; Choi et al. 2010).

\section{Systematic Effects and Modeling}
\subsection{Finite pixel size}\label{Sec:pix}

The density field estimated from a discrete particle distribution
and mapped onto a grid is affected by the finite pixel size and 
the way that the particle mass is divided into adjacent pixels.
To study the effects as a function of pixel size we use a set of grids 
with cubic pixels and with size of
600, 720, 840, 900, 960, 1080, 1200, 1260, 1440, 1800, 1920, and 2160
in terms of number of pixels along a side.
Since the physical size of HR2 is 7200 $\hMpc$, they correspond to
12, 10, 8.57, 8, 7.5, 6.67, 6, 5.71, 5, 4, 3.75, and $3.33\hMpc$
respectively, in terms of pixel size $p$.
We used the Cloud-in-Cloud (CIC; Hockney \& Eastwood 1981)
mass assignment scheme in this study. 
The pixel densities are then smoothed over 
three different smoothing lengths, $R_G$ = 15, 22, and $34\Mpc$.
These smoothing scales are motivated by the mean separations of galaxies in
the SDSS-III BOSS CMASS sample ($\bar{d}=15\hMpc$) and SDSS-II LRG sample 
(Gott et al. 2009). 
The genus curves are obtained for each case of selected grid size and smoothing 
scale.
Also we calculate the mean-square matter density $\sigma_0$ and 
the mean square density gradient fluctuations $\sigma_1$.


\begin{figure}
\epsscale{1}
\plotone{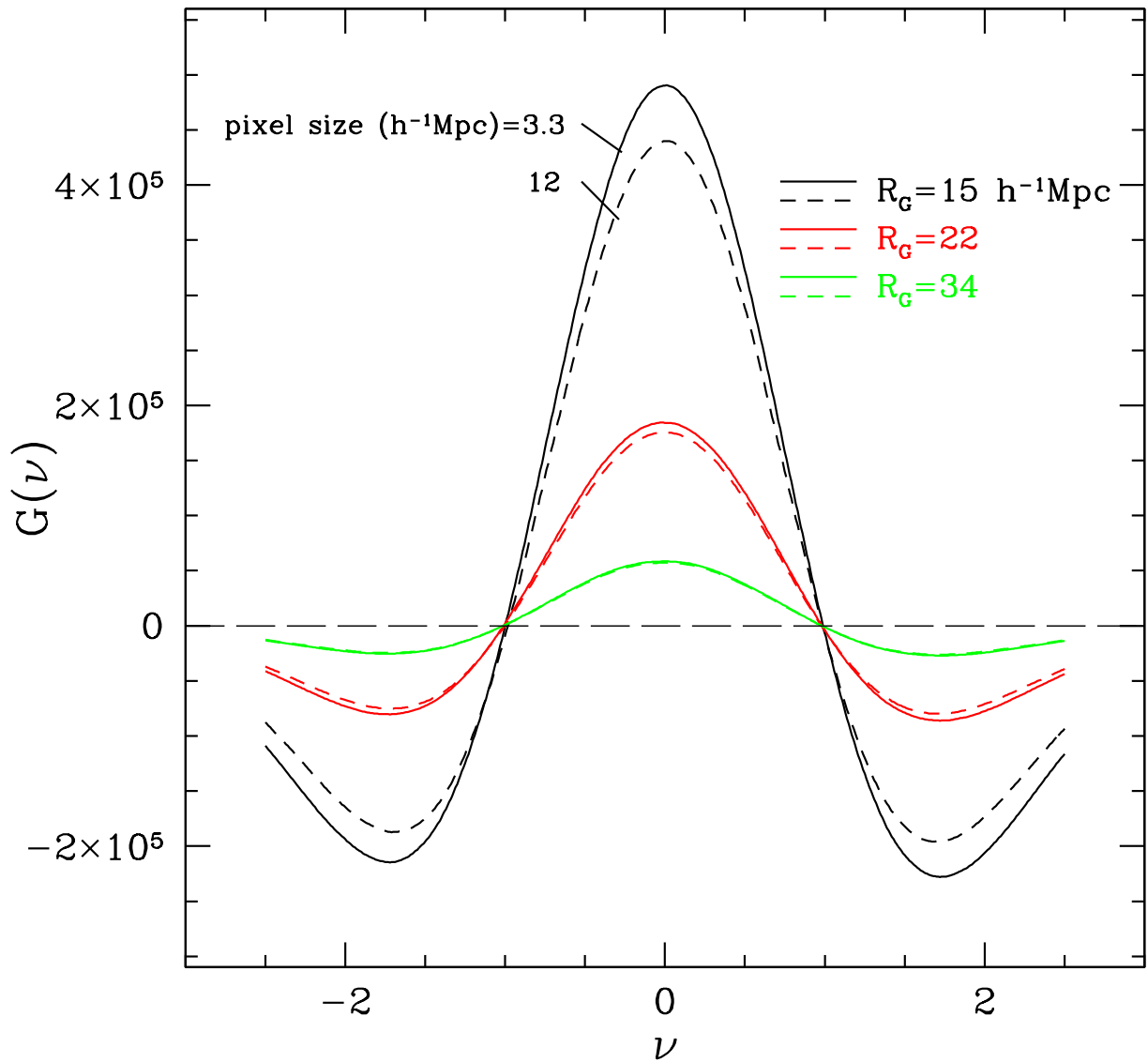}
\caption{
Pixel-size dependence of the genus curve on three smoothing scales.
The solid (dashed) lines are the genus curve obtained with the grids of size 
$p=3.3 (12)\hMpc$.}
\label{Fig:gpix}
\end{figure}

\begin{figure}
\epsscale{1}
\plotone{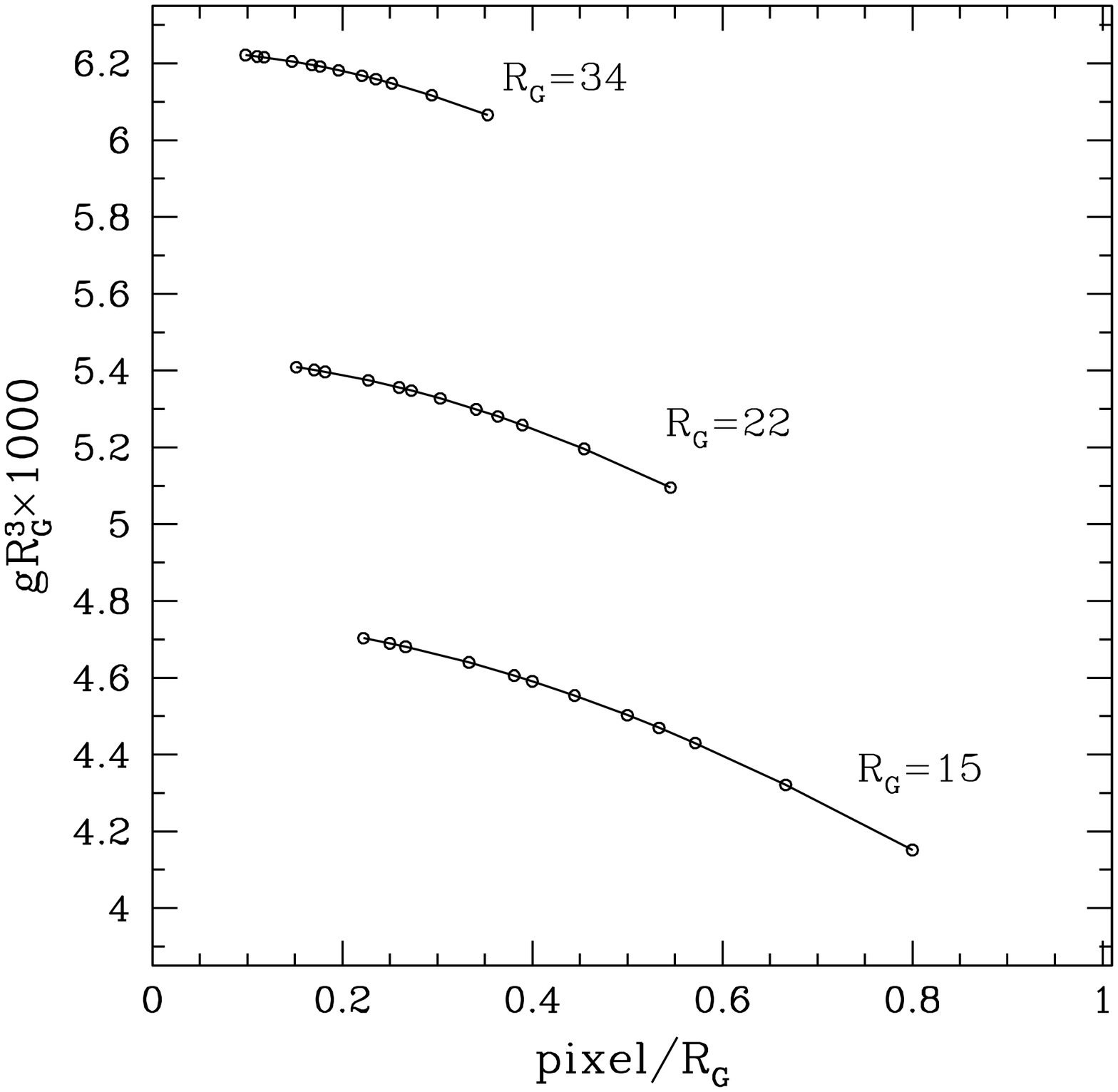}
\caption{
Dependence of the amplitude of the genus curve on the relative pixel size.
The amplitude is calculated from Equation~(\ref{Eq:gamp}).
}
\label{Fig:pixgamp}
\end{figure}

Figures~\ref{Fig:gpix} and \ref{Fig:pixgamp} show how the amplitude of the genus curves depends on 
the pixel size relative to the smoothing length.
In Figure~\ref{Fig:pixgamp} the amplitude $A$ is calculated for each $p/R_G$ using
the formula for Gaussian fields (Gott et al. 1986; Tomita 1986)
\begin{equation}
A =\frac{1}{(2\pi)^{2}} \left[\frac{\sigma_{1}}{\sqrt{3}\sigma_{0}}\right]^{3}, 
\label{Eq:gamp}
\end{equation}
where $\sigma_0$ and $\sigma_1$ are directly calculated from the simulation.
In many previous analyses of observational and simulation data a rule of thumb
for choosing the pixel size was $p=\sqrt{2} R_G /2.5$ or $p/R_G=0.566$ (Gott et al. 1989).
As can be seen in Figure~\ref{Fig:pixgamp}, such a choice results in underestimation of the genus
amplitude by more than 5\%.

\begin{table*}
\begin{center}\label{Tab:fitpix}
\caption{Coefficients for pixel effects
}
\centering\tiny
\begin{tabular}{r|rrrr| rrrr| rrrr}
\hline \hline
\multicolumn{13}{l}{$\Delta G_{\rm pixel}/A=e^{-\nu^2/2}[aH_0+bH_1(\nu)+cH_2(\nu)+dH_4(\nu)]p^2/R_G^2$}\\
\hline
& \multicolumn{4}{c|}{$R_G =15~h^{-1}\rm Mpc$} & \multicolumn{4}{c|}{$R_G =22~h^{-1}\rm Mpc$} &
\multicolumn{4}{c}{$R_G=34~h^{-1}\rm Mpc$}\\
\cline{2-5} \cline{6-9}\cline{10-13} 
$p$ & $a$ & $b$ & $c$ & $d$ & $a$ & $b$ & $c$ & $d$ & $a$ & $b$ & $c$ & $d$\\
\hline
3.33 &  0.0347 &  0.0197 &  0.2866 & 0.0294 & 0.0446 & 0.0228 & 0.3300 & 0.0377& 0.0515 & 0.0175 & 0.3597 & 0.0454\\
3.75 &  0.0390 &  0.0214 &  0.2982 & 0.0313 & 0.0481 & 0.0235 & 0.3389 & 0.0395& 0.0527 & 0.0194 & 0.3637 & 0.0464\\
4.00 &  0.0429 &  0.0230 &  0.3054 & 0.0325 & 0.0501 & 0.0232 & 0.3405 & 0.0397& 0.0562 & 0.0185 & 0.3642 & 0.0457\\
5.00 &  0.0440 &  0.0234 &  0.3076 & 0.0329 & 0.0450 & 0.0248 & 0.3410 & 0.0350& 0.0539 & 0.0192 & 0.3633 & 0.0462\\
5.71 &  0.0441 &  0.0234 &  0.3089 & 0.0334 & 0.0511 & 0.0246 & 0.3436 & 0.0404& 0.0538 & 0.0201 & 0.3607 & 0.0451\\
6.00 &  0.0454 &  0.0244 &  0.3102 & 0.0335 & 0.0504 & 0.0234 & 0.3398 & 0.0397& 0.0534 & 0.0190 & 0.3577 & 0.0444\\
6.67 &  0.0463 &  0.0239 &  0.3105 & 0.0334 & 0.0512 & 0.0268 & 0.3423 & 0.0401& 0.0546 & 0.0200 & 0.3605 & 0.0462\\
7.50 &  0.0465 &  0.0231 &  0.3111 & 0.0335 & 0.0508 & 0.0244 & 0.3358 & 0.0382& 0.0548 & 0.0166 & 0.3625 & 0.0461\\
8.00 &  0.0457 &  0.0238 &  0.3094 & 0.0336 & 0.0507 & 0.0246 & 0.3415 & 0.0403& 0.0538 & 0.0201 & 0.3579 & 0.0441\\
8.57 &  0.0447 &  0.0240 &  0.3054 & 0.0328 & 0.0499 & 0.0248 & 0.3391 & 0.0397& 0.0496 & 0.0229 & 0.3481 & 0.0427\\
10.00 &  0.0432 &  0.0228 &  0.3021 & 0.0325 & 0.0490 & 0.0235 & 0.3357 & 0.0394& 0.0527 & 0.0181 & 0.3506 & 0.0444\\
average: &  0.0433 &  0.0230 &  0.3050 & 0.0326 & 0.0496 & 0.0242 & 0.3389 & 0.0395& 0.0534 & 0.0192 & 0.3590 & 0.0451\\
\hline
\end{tabular}\end{center}
\tablecomments{Pixel size $p$ is in units of $\hMpc$. Amplitude $A$ is 524268, 190332, and 59180 for $R_G=15$, 22, and 
$34\hMpc$, respectively.}
\end{table*}

\begin{figure*}
\epsscale{1}
\plotone{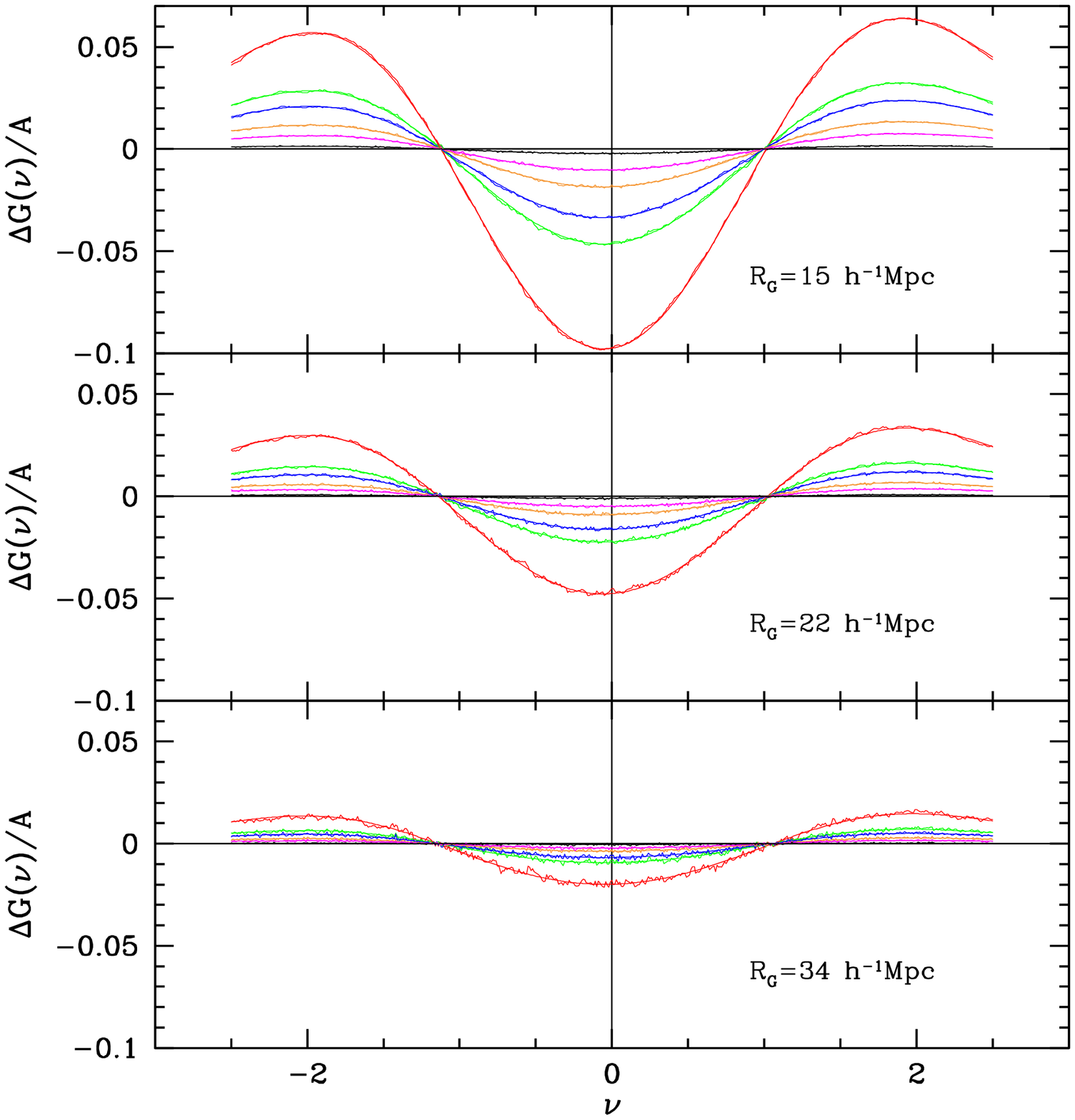}
\caption{
Deviations of the genus curve from the Gaussian value due to
the finite pixel size effects.
Each color denotes the pixel
size - 3.33 (black), 5(magenta), 6(yellow), 7.5(blue), 8.57(green) and 12(red) $h^{-1}\rm Mpc$.
\label{Fig:delgpix}
}
\end{figure*}

In this section we aim to calculate the genus curves $G(\nu; p/R_G)$ and
$G(\nu; p/R_G=0)$ with and without the finite pixel size effects, respectively.
Their difference is the systematic effects, which we model by the formula
\begin{eqnarray}
\nonumber
&&\Delta G_{\rm pixel}(\nu)=A e^{-\nu^2/2} \times\\
&&[aH_0(\nu)+bH_1(\nu)+cH_2(\nu)+dH_4(\nu)]p^2/R_G^2,
\label{Eq:delgpix}
\end{eqnarray}
where $H_i$'s are the Hermite polynomials\footnote{
$H_i$'s are the Hermite polynomials, and $H_0(\nu)=1$,
$H_1(\nu)=\nu$, $H_2(\nu)=\nu^2-1$, $H_3(\nu)=\nu^3-3\nu$, and
$H_4(\nu)=\nu^4-6\nu^2+3$.}
and
$A$ is the amplitude of the genus curve of a Gaussian field having the same 
$\sigma_0$ and $\sigma_1$ measured with vanishing size pixels.

We find that this functional form very accurately model the pixel effects.  
This form is partly motivated by Hamilton et al. (1986), who predicted 
the odd function term $bH_1$ to be absent.

Since we do not know the genus amplitude with no pixel effects in the beginning,
we cannot calculate the difference $\Delta G$ directly.
We therefore obtain the genus amplitude and the coefficients at the same time iteratively. 
At the first step we calculate $\Delta G$ and the coefficients assuming that
the genus amplitude calculated according to Equation~(\ref{Eq:gamp}) from the density 
field constructed with the highest resolution approximates the genus 
with no pixel effects. In the next step
the new amplitude is estimated using Equations~(\ref{Eq:gamp}) and (\ref{Eq:delgpix}) 
from $A(p/R_G=0) = A(p_s /R_G)-\Delta G(\nu=0; p_s /R_G)$,
where $p_s$ is the smallest pixel size. The updated amplitude $A$ is
524268, 190332, and 59180 for $R_G=15$, 22, and $34\hMpc$, respectively.
From the updated amplitude the new $\Delta G$ and coefficients are obtained.

To find the coefficients $a$, $b$, $c$, and $d$ we carried out a chi-square 
($\chi^2$) minimization with Markov Chain Monte Carlo (MCMC) method
by following the steps described in Verde et al. (2003) over ranges $-2.5<\nu<2.5$.
We measure the coefficients for a set of values of $p/R_G$. 
For example, in the case of $R_G=15\hMpc$ we measure them for 
$p/R_G=$ 0.222, 0.25, 0.267, 0.333, 0.381, 0.4, 0.445, 0.5, 0.533, 0.571, 0.667, 
and 0.8. 
The smallest value corresponds to the biggest mesh $2160^3$, and 
the largest one to the smallest mesh $600^3$.
The iterative process makes the coefficients independent of
the pixel size as assumed in Equation~(\ref{Eq:delgpix}).
The set of coefficient values that minimize $\chi^2$ are listed in 
Table~\ref{Tab:fitpix}
for three smoothing lengths.
In Table~\ref{Tab:pix} in Appendix~\ref{App:genus}, we list the complete genus
values relative to these measurements as a function of the volume
fraction threshold $\nu$. 

In Figure~\ref{Fig:delgpix} we show the finite pixel size effects normalized by 
the genus amplitude without pixel effect.
Shown are cases for $p=3.33$, 5, 6, 7.5, 8.75, and $12\hMpc$.
The smooth curves in the same color are the fits given by Equation~(\ref{Eq:delgpix})
with the coefficients listed in Table~\ref{Tab:fitpix}. 
The fits are so excellent that it is difficult to distinguish between 
the numerical data and the corresponding fit. 
We dropped the term $H_3$ in the final form of the model because we found its 
contribution is negligible."

It can be seen in Figure 3 that if one follows Park et al. (2005), 
who argued that the pixel size should be smaller than
the smoothing length at least by a factor three, the underestimation of
the genus amplitude becomes less than 1\%.

Hamilton et al. (1986) derived a formula for the pixel effects that includes
the even Hermite polynomials such as $H_0$, $H_2$ and $H_4$.  But we find 
that the genus deviations are not exactly symmetric with respect to $\nu=0$
and an odd term ($H_1$) is needed to better describe the pixel effects.
This is because the high-density regions are more affected by the pixel smoothing.
But it is still true that the even terms are dominant and
the consequence brought by pixel effects on the genus curve is 
mostly the amplitude drop. 

\subsection{Non-linear gravitational evolution}\label{Sec:grav}

The genus of the large-scale density field is insensitive to the
non-linear gravitational evolution because it remains the same
as far as the connectivity of structures does not change, regardless
of change in the amplitude or shape of the fluctuations.
This makes the genus a better tool for a cosmological probe than 
galaxy power spectrum or correlation function where it is not so 
easy to model and remove the large non-linear effects 
(see Figs. 4-6 and 11 of Kim et al. 2009).
The deviation of the genus curve from its Gaussian form has been 
studied in many literatures (Melott et al. 1988; Park \& Gott 1991;
Weinberg \& Cole 1992; Matsubara 1994, 2003;
Park et al. 2005; Hikage et al. 2003, 2006), where most of the analytic
derivations are based on the second-order perturbation theory.

To estimate the non-linear gravitational evolution effects on 
the genus deviations, we measure the genus curves of
the initial matter density field (at $z=32$) and 
of the final matter density field (at $z=0$) 
using a large array of $2160^3$ pixels to minimize the pixel effects.
In Table~\ref{Tab:grav} in Appendix~\ref{App:genus},
the genus values are given as a function of the volume fraction threshold,
$\nu$.

\begin{figure}
\epsscale{1.}
\plotone{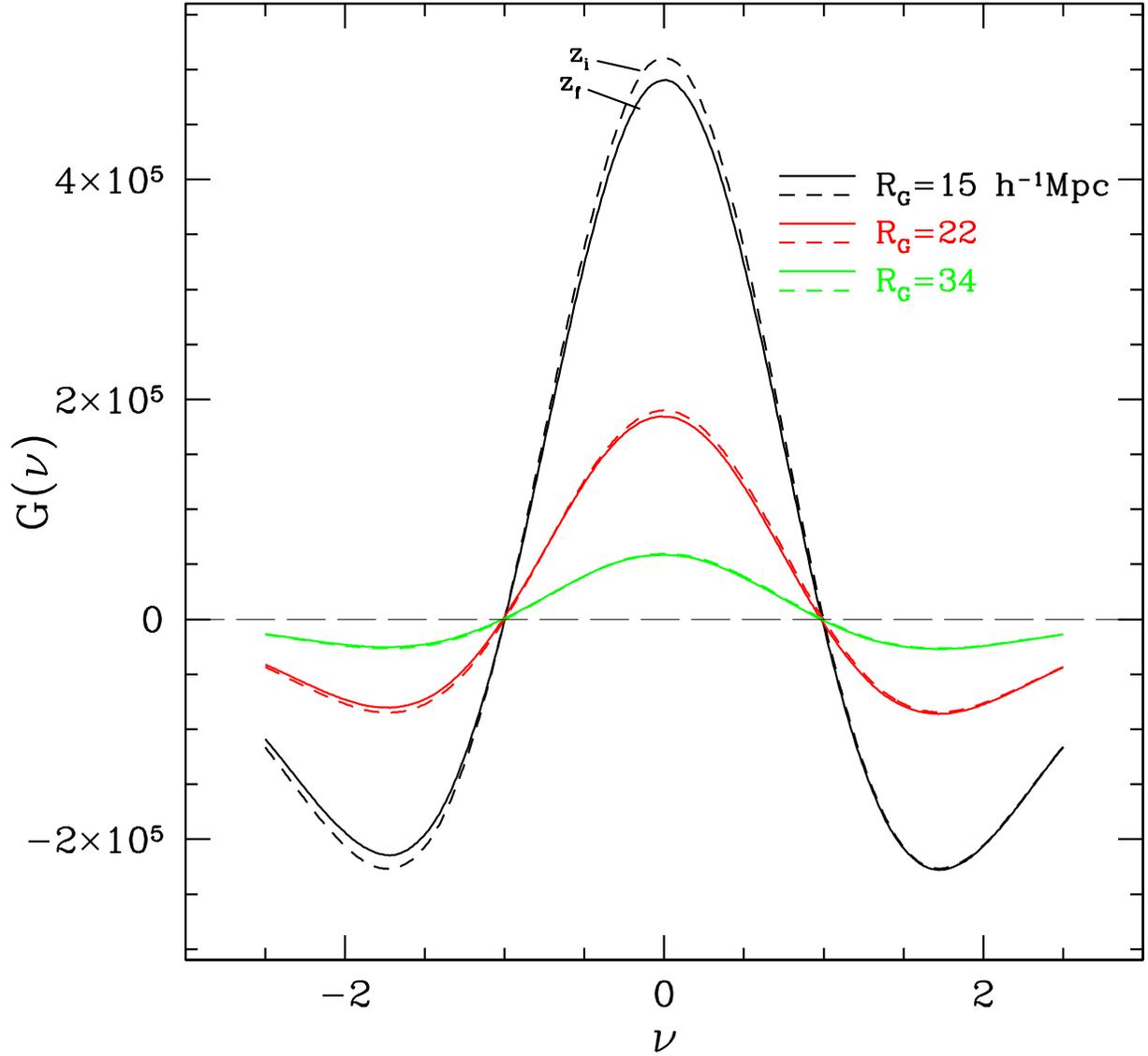}
\caption{
Gravitational evolution effects on the genus curve. The solid and
dashed curves are the genus measured from the density fields at the initial 
($z_{\rm i}=32$) and final ($z_{\rm f}=0$) epochs, respectively.}
\label{Fig:ggrav}
\end{figure}

\begin{figure*}
\epsscale{1}
\plotone{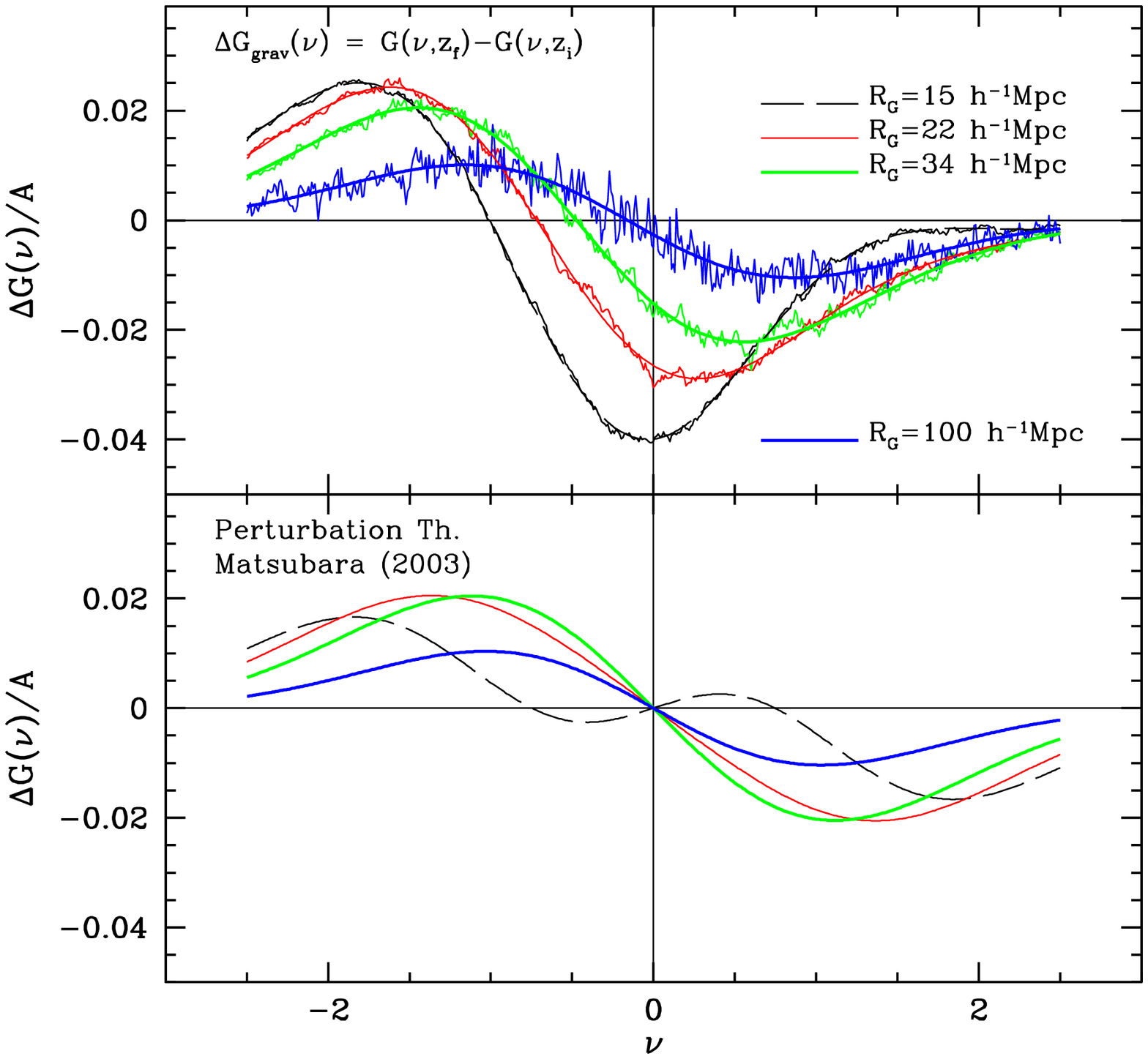}
\caption{
Deviations of the genus curve from the Gaussian function due to the 
gravitational evolution at four smoothing scales.
Curves in the upper panel are from simulation,
and those in the lower panel are predictions by the perturbation theory.
In the top panel, the curves with jitters are the numerically measured 
genus deviations and the smooth curves are fitting functions 
using the Hermite polynomials up to the 4th order.
At $R_G=100\hMpc$, the perturbation theory well agrees with the simulation.
}
\label{Fig:delggrav}
\end{figure*}

Figure~\ref{Fig:ggrav} shows the genus curves of the initial and
final matter density fields, which clearly show 
amplitude drop due to the nonlinear gravitational evolution.
The gravitational evolution affects the 
topology of low-density regions more seriously.
This is because, as the gravitational evolution progresses,
expanding low-density regions percolate with one another
by opening their outer boundaries,
and the number of disconnected voids decreases.
On the other hand, the high-density regions contract 
to remain as density peaks surrounded by less dense regions
unless they happen to have very close merging pairs.



We compute the difference of the genus ($\Delta G_{\rm grav}$) 
between the initial matter density 
field at $z_{\rm i}=32$ and the final matter density field at $z_{\rm f}=0$ as
\begin{equation}
\Delta G_{\rm grav}(\nu,z,R_G)=G(\nu,z_{\rm f},R_G)-G(\nu,z_{\rm i},R_G),
\label{Eq:delgsim}
\end{equation}
and normalize it with the genus amplitude using Equation~(\ref{Eq:gamp}).
Since all other conditions are fixed, this  will give us 
the sole effect of non-linear gravitational evolution.

Matsubara (1994, 2003)
derived an explicit formula for the genus due to the gravitational
evolution including 
nonlinear terms from the second order perturbation theory.
For three-dimensional genus and when the volume-fraction thresholds
are used, this has the form of
\begin{eqnarray}
\nonumber
&&G(\nu)=-A e^{-\nu^2/2} \times \\ 
\nonumber
&&\left[H_2(\nu)
 + [(S^{(1)}-S^{(0)})H_3(\nu)+(S^{(2)}-S^{(0)})H_1(\nu)]\sigma_0 \right]\\
\label{Eq:gpert}
\end{eqnarray}
where $S^{(i)}$ are the skewness parameters (see Matsubara 2003 for their definitions).

Figure~\ref{Fig:delggrav} shows the difference calculated from the simulation 
results (top panel) and that from the analytic prediction 
by Matsubara (bottom panel).
In Table~\ref{Tab:fitgrav}, 
we summarize the values of  $A$, $\sigma_0$, $\sigma_1$, 
and fitting coefficients.
According to the Matsubara's formula, the genus deviations should be 
symmetric with respect to the origin with opposite sign, and at $\nu=0$ 
the genus deviation is expected to be null. On the other hand,
the simulation measurement shows that the deviation is far from 
symmetric and there is a significant drop at $\nu=0$.

We note that the measured $\Delta G_{\rm grav}$ is
significantly shifted towards the low density regions.
The amplitude drop implies that there need terms with even orders ($H_{2n-2}$).
We model $\Delta G_{\rm grav}$ as follows:
\begin{eqnarray}
\nonumber
&&\Delta G_{\rm grav}(\nu)=A e^{-\nu^2/2}\times \\
&&[(bH_1(\nu)+dH_3(\nu))\sigma_0+(aH_0(\nu)+cH_2(\nu))\sigma_0^2].
\label{Eq:delggrav}
\end{eqnarray}
$A$ is again the genus amplitude estimated from the Gaussian formula 
(Eq.~\ref{Eq:gamp}), where $\sigma_0$ and $\sigma_1$ are directly measured 
from the initial density field and linearly extrapolated to $z=0$ 
(rather than the final density field).
In this equation, the first two terms in the square bracket 
can be also found in Equation~(\ref{Eq:gpert}) and the other terms with 
even functions
are added to model the rest of deviations due to high-order non-linearities.
In Table~\ref{Tab:fitgrav} we show that as the smoothing length increases,
the coefficients of the Hermite polynomials calculated from the 
perturbation theory, $S^{0}-S^{1}$ and $S^{0}-S^{2}$ become in good agreement 
with the coefficients $b$ and $d$ in Equation~(\ref{Eq:delggrav}), respectively.
In other words, the significant discrepancy between the gravitational effects
found in simulation and the Matsubara's prediction is mainly due to the lack of
the even functions in the analytic formula.
The $\rms$ density fluctuation on $R_G=15, 22,$ and 34$h^{-1}$Mpc scales 
is 0.2604, 0.1763, and 0.1080, respectively (see Tab.~\ref{Tab:fitgrav}). 
The agreement becomes much better for $R_G=100h^{-1}$Mpc when $\sigma_0=0.0243$ 
(see the blue curves in Fig.~\ref{Fig:delggrav}).

\begin{figure}
\epsscale{1}
\plotone{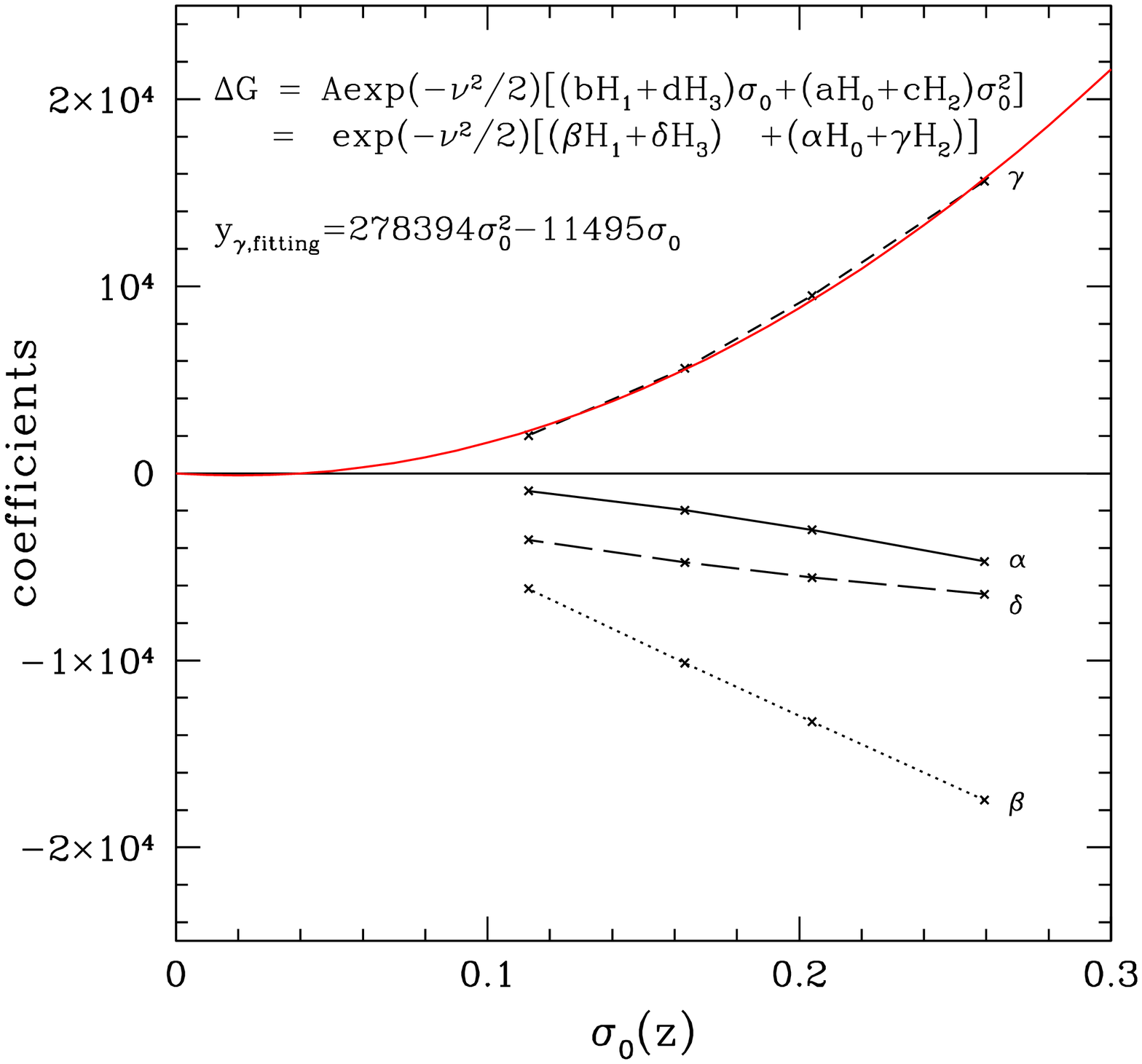}
\caption{
Change of the fitting coefficients with the standard deviation
of the matter density field. Here, we use the definitions,
$\alpha \equiv Aa\sigma_0^2, \beta \equiv Ab\sigma_0,
\gamma \equiv Ac\sigma_0^2,
\delta \equiv Ad\sigma_0$
}
\label{Fig:fitgrav}
\end{figure}

We overplot the best fit model (smooth curves) in the top panel 
of Figure~\ref{Fig:delggrav} to demonstrate that our model describes the gravitational
evolution much more accurately than Matsubara's.
Since our simulated genus is different from the analytical prediction,
we take one more step to cross check the dependence of
the genus on  $\sigma_0$ in Equation~(\ref{Eq:delggrav}).
%
We use the HR2 simulation cube data at four redshifts ($z$ = 2, 1, 0.5,
and 0) to measure $\sigma_0$ and the genus, and fit the genus with
Equation~(\ref{Eq:delggrav}).
The behavior of the coefficients of the Hermite polynomials
is shown in Figure~\ref{Fig:fitgrav}.
 During the measurement, we use the simplified coefficients
 such as $\alpha\equiv Aa\sigma_0^2$ and $\gamma \equiv Ac\sigma_0^2$.
 The figure shows that the coefficients for $H_0$ and $H_2$
 have quadratic dependence on $\sigma_0$, validating our fitting model.
 This means that we need not only a linear order term in $\sigma_0$ but also
 the second-order terms.
Therefore, adding one more term to the perturbation theory
 up to the second order in $\sigma_0$
 will produce a significantly better fit to the simulated data.
%
\begin{table}
\begin{center}
\caption{Coefficients for gravitational evolution effects for the matter density fields}
\begin{tabular}{llllllll}
\hline
\hline
\multicolumn{8}{l}{$\Delta G_{\rm grav}/A=e^{-\nu^2/2}[(bH_1+dH_3)\sigma_0+(aH_0+cH_2)\sigma_0^2]$} \\
$R_G$ & $a$  & $b$  & $c$  & $d$ & $\Delta_{\rms}$ &$\sigma_0$ & $\sigma_1$\\
\hline
  15 & -0.1365 & -0.1317 & 0.4523 & -0.0487 & 3.9E-4 & 0.26039 & 0.01703 \\ 
  22 & -0.1750 & -0.2310 & 0.6779 & -0.0433 & 6.9E-4 & 0.17634 & 0.00830 \\
  34 & -0.2247 & -0.3334 & 1.0791 & -0.0336 & 7.4E-4 & 0.10797 & 0.00345 \\
 100 & -0.6263 & -0.7127 & 3.8696 & -0.0151 & 8.1E-4 & 0.02429 & 0.00031 \\
\hline
\hline
\multicolumn{8}{l}
{$\Delta G_{\rm grav}^{\rm PT}/A_{\rm LT}=e^{-\nu^2/2}[(S^{(0)}-S^{(2)})H_1+(S^{(0)}-S^{(1)})H_3]\sigma_0$} \\
$R_G$ & $S^{(0)}-S^{(2)}$ & $S^{(0)}-S^{(1)}$ & $\sigma_0$ & $\sigma_1$& \\
\hline
 15 & -0.1616 & -0.0662 & 0.26225 & 0.01727 \\
 22 & -0.2732 & -0.0501 & 0.17665 & 0.00837 \\
 34 & -0.3727 & -0.0313 & 0.10759 & 0.00345 \\
100 & -0.7494 & -0.0212 & 0.02418 & 0.00031 \\
\hline
\end{tabular}
\label{Tab:fitgrav}
\end{center}
\tablecomments{
Smoothing length $R_G$ is in units of $\hMpc$. 
The smallest pixel size of $p=3.33\hMpc$ is used
to minimize the pixel effect. Amplitude $A$, directly measured from 
the matter density field is 509401, 189590, 59366, and 3791 for $R_G=15$, 22, 34, and
$100\hMpc$, respectively. $\Delta_{\rms}$ is calculated from
$\sqrt{\sum\limits_{j=1}^{N} \frac{1}{N} [\Delta G_{\rm grav}^{\rm sim}(\nu_j)/A 
-\Delta G_{\rm grav}(\nu_j)/A]^2}$ 
where $\Delta G_{\rm grav}^{\rm sim}(\nu_j)$ is from simulation (Eq.~\ref{Eq:delgsim})
and $\Delta G_{\rm grav}(\nu_j)$ is from the best-fit function (Eq.~\ref{Eq:delggrav}).
Amplitude $A_{\rm LT}$ is the prediction from linear theory.
}
\end{table}

\subsection{Redshift-space distortion}\label{Sec:RSD}


In order to quantify the redshift-space distortion (RSD) effects 
on the genus curve,
we compute the genus of the matter and halo density fields in 
real and redshift spaces. We again use the smallest possible pixel size 
($2160^3$ mesh for matter and $2100^3$ mesh for halos) to 
minimize any unwanted other systematic effects.
In Table~\ref{Tab:RSD} in Appendix~\ref{App:genus}, the genus curves 
are given as a function of $\nu$.

\begin{figure}
\epsscale{1}
\plotone{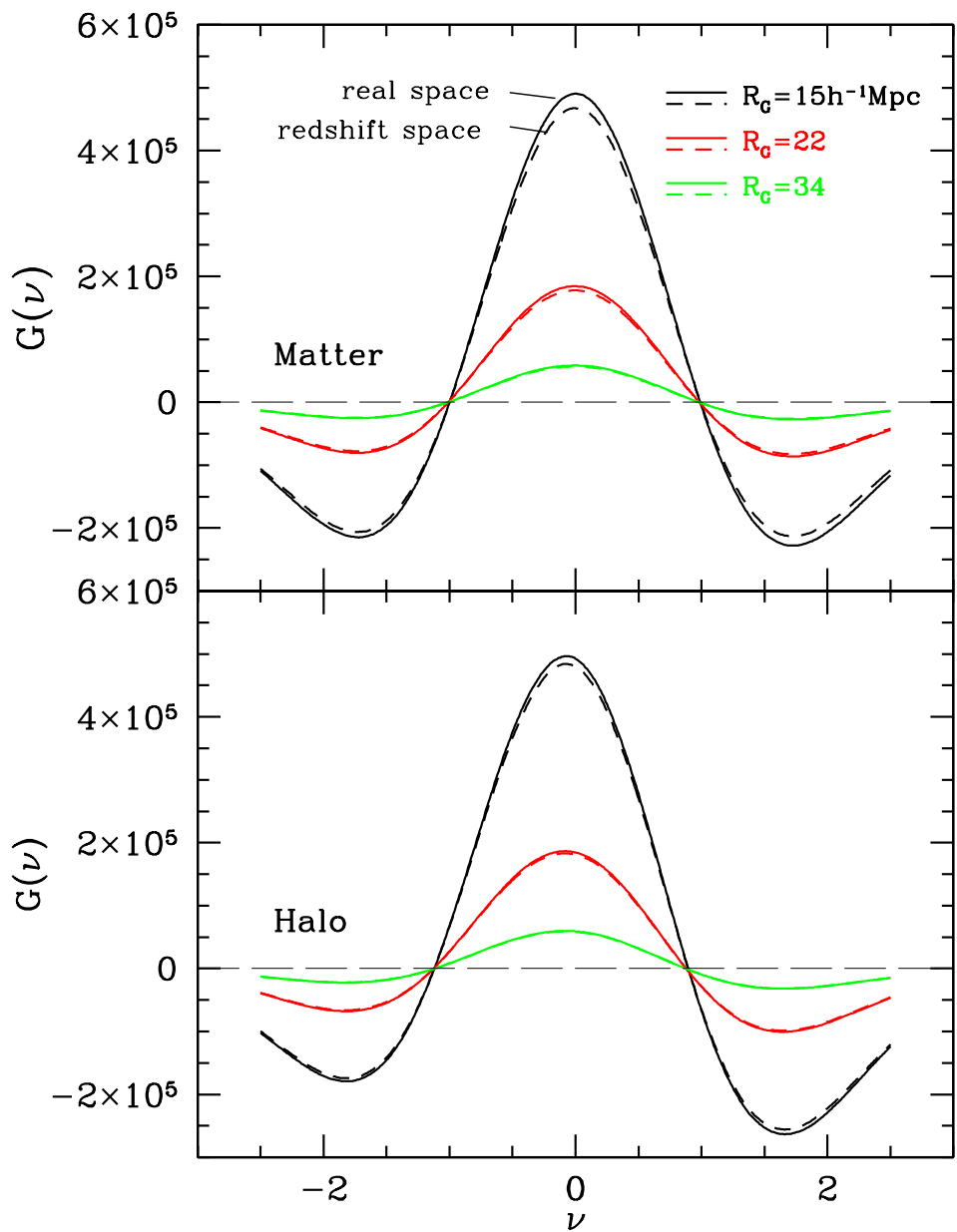}
\caption{
Genus curves in redshift (dashed) and real (solid lines) spaces for
matter (top) and halo (bottom panel) density fields at three
smoothing scales, $R_G$ = 15 (black), 22 (red), and $34\hMpc$ (green). 
The redshift distortion reduces the genus
amplitude in both matter and halo samples. 
Halo is less susceptible to redshift distortion.
}
\label{Fig:gRSD}
\end{figure}

\begin{figure*}
\epsscale{1}
\plotone{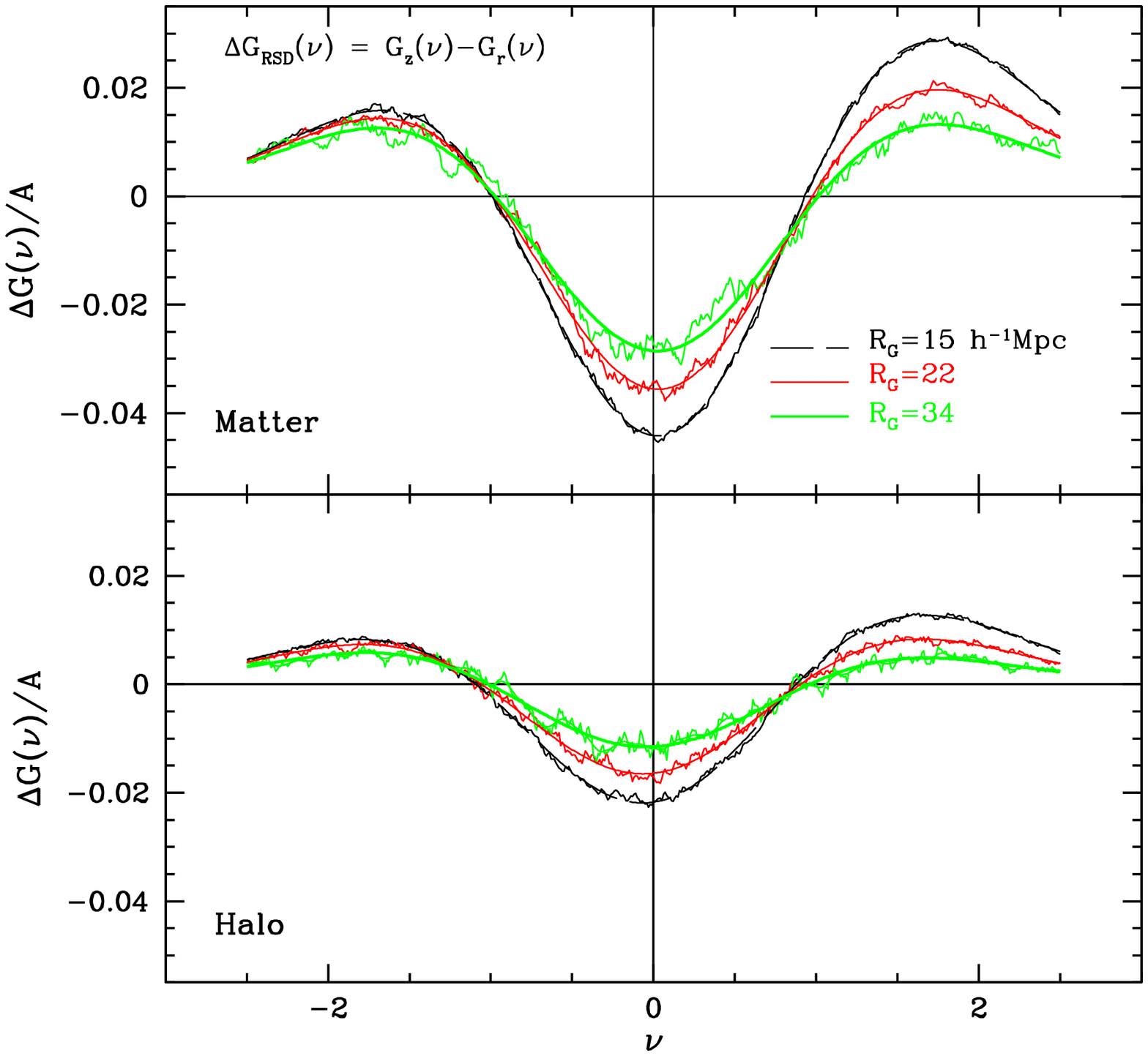}
\caption{
Effects of the redshift-space distortion on the genus in the matter density
field (top) and in the halo density field (bottom).
Y-axis is the normalized genus deviations caused by the redshift-space 
distortion. The jittering lines are obtained from simulation 
and smooth curves are our best fit models.
}
\label{Fig:delgRSD}
\end{figure*}

Matsubara (1996) derived an analytical expression for 
the RSD effects on the genus curve in the linear regime as follows:
\begin{equation}
G_{z}(\nu)=\frac{3\sqrt{3}}{2}\sqrt{C}(1-C)G_{r}(\nu),
\end{equation}
where $G_{z}(\nu)$ and $G_r(\nu)$ are the genus curves
in redshift space and real space, respectively, and $C\equiv C_1/C_0$. 
According to Matsubara (1996), 
\begin{equation}
C_j \equiv {1\over2}\int_{-1}^1 d\mu \mu^{2j} (1+fb^{-1} \mu^2)^2,
\end{equation}
where $b$ is the constant bias factor, $f\equiv d\ln D/d\ln a$,
$D$ is the growth factor, and $a$ is the cosmic expansion parameter.
This formula suggests that the RSD does not change 
the shape of the genus curve but reduces its amplitude.
This linear theory prediction does not agree well with our results 
on the smoothing scales we study as shown below.

In Figure \ref{Fig:gRSD} we plotted
genus curves in real (solid) and redshift (dashed lines) spaces.
In the top panel, we plotted genus curves for matter density field at $z=0$.
As one can see, the genus amplitude in the high density regions increases slightly 
more than in the low density regions.
The halo density field shows a smaller amplitude drop than the matter density field,
which implies that halos, being the massive objects, are 
less susceptible to RSD than matter.

\begin{table}
\begin{center}
\label{Tab:fitRSD}
\caption{Coefficients for redshift-space distortion effects for the matter density fields}
\begin{tabular}{llllllll}
\hline
\hline
\multicolumn{8}{l}{$\Delta G_{\rm RSD}/A=e^{-\nu^2/2}(aH_0+bH_1+cH_2+dH_3)$}\\
$R_G$ & $a$ & $b$ & $c$ & $d$ & $\Delta_{\rms}$ & $\sigma_0$ & $\sigma_1$  \\
\hline
  15 &0.0037  &  0.01677  & 0.0478 & 0.0068 & 3.6e-4 & 0.25947  & 0.01709 \\
  22 &0.0017  &  0.00698  & 0.0372 & 0.0034 & 7.9e-4 & 0.17529  & 0.00825 \\
  34 &0.0003  &  0.00098  & 0.0288 & 0.0010 & 15.2e-4& 0.10745  & 0.00343 \\
\hline
\end{tabular}
\end{center}
\tablecomments{Smoothing length $R_G$ is in units of $\hMpc$. 
The smallest pixel size of $p=3.33\hMpc$ is used to minimize the pixel effect. 
Amplitude $A$ is 520141, 189599, and 59084 for $R_G=15$, 22, and
$34\hMpc$, respectively.
$\Delta_{\rms}$ is calculated from
$\sqrt{\sum\limits_{j=1}^{N}\frac{1}{N}
[\Delta G_{\rm RSD}^{\rm sim}(\nu_j)/A-\Delta G_{\rm RSD}(\nu_j)/A]^2}$
where $\Delta G_{\rm RSD}^{\rm sim}(\nu_j)$ is from simulation 
and $\Delta G_{\rm RSD}(\nu)$ is the best-fit function.
}
\end{table}             
                         
With the definition of 
$\Delta G_{\rm RSD} \equiv G_{\rm z}-G_{\rm r}$, 
we quantify the effects of RSD and fit the results with the fitting function 
\begin{eqnarray}
\nonumber
\Delta G_{\rm RSD}(\nu)&&=A e^{-\nu^2/2} \times \\
&&[aH_0(\nu)+bH_1(\nu)+cH_2(\nu)+dH_3(\nu)],
\end{eqnarray}
where the amplitude $A$ is estimated using the Gaussian formula
for the matter (or halo) density fields in real space.
The change in the genus 
is well-modeled with a combination of 
Hermite polynomials up to the third order (see top panel of Fig.~\ref{Fig:delgRSD}
for the matter density field and
bottom panel for the halo density field).
The values of coefficients are listed in Table~\ref{Tab:fitRSD}.

It can be seen in these figures that the genus in the high density regions
decreases more compared to that in low density regions. 
This clearly indicates that
the asymmetry is produced by the excessive mergers of high-density regions
along the line of sight directions.
The odd terms ($H_1$ and $H_3$) in the model explain the asymmetry.
Although the coefficients $b$ and $d$ of the $H_1$ and $H_3$ 
for $R_G=15$ and $22\hMpc$ listed in Table~\ref{Tab:fitRSD} are very small 
values, they are significantly non-zero.
The asymmetry disappears rapidly as the smoothing length increases.

\subsection{Shot noise effects and halo biasing}\label{Sec:shot}

Discrete sampling of an underlying density field at finite
number of points produces the shot noise effects on the genus.
%
The situation is more complicated when we study 
the halo (or galaxy) density field because the shot noise and 
the halo biasing effects are entangled.
Separating these two effects is hard because the halos with mass
above a given value is intrinsically sparse and we cannot avoid studying
these effects in combination.


\begin{figure}
\epsscale{1}
\plotone{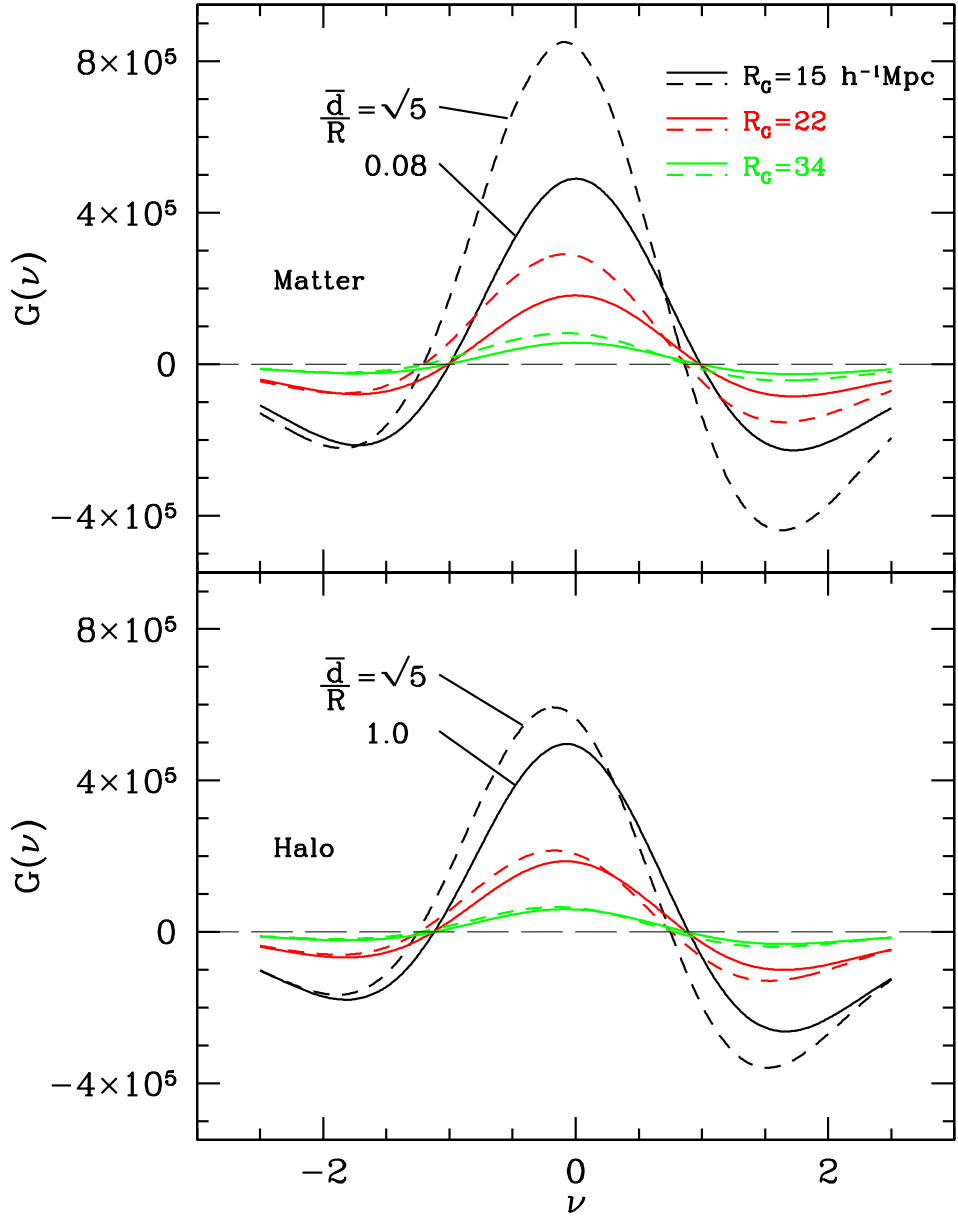}
\caption{
Effects of shot noise in the matter density field (top) and
in the halo density field (bottom).
Dashed lines are the genus for $\bar{d}/R_G$ = $\sqrt{5}$, and solid lines are for $\bar{d}/R_G$ = 
0.08 (matter) and 1 (halo). Black, red, and green curves are obtained using $R_G$ = 15, 22, and 
$34\hMpc$, respectively. 
The shot noise not only changes the genus amplitude but also shifts the genus curve to the low density 
region, which breaks the symmetry of the genus curve.
}
\label{Fig:gshot}
\end{figure}




First, we study the pure shot noise effects in dark matter distribution.
For each of three smoothing lengths, 
we have randomly sampled simulation particles at z = 0 to have mean particle
separations of
$\bar{d}/R_G=$0.08, 0.16, 0.32, 0.64, 1.0, $\sqrt{2}$, $\sqrt{3}$, 2.0, and $\sqrt{5}$.
In this case, we used the smallest pixel size (or the mesh size of $2160^3$).
In Table~\ref{Tab:shotm} in Appendix~\ref{App:genus}, the genus values 
are given as a function of $\nu$.
In the top panel of Figure~\ref{Fig:gshot}, we plot genus curves for the smallest (solid lines) and the biggest $d/R_G$ (dashed lines). 
As $d/R_G$ increases, the genus curve
shifts to the low density regions and  
the amplitude changes differently in low and high density regions.
High density regions suffer from shot noise much more than low density regions.
This is because some structures become split into 
multiple objects as we take out dark matter particles in the 
high density region, resulting in more isolated regions.
The distortion of the genus curve is much larger
compared to all the other systematic effects discussed so far.

\begin{figure*}
\epsscale{1}
\plotone{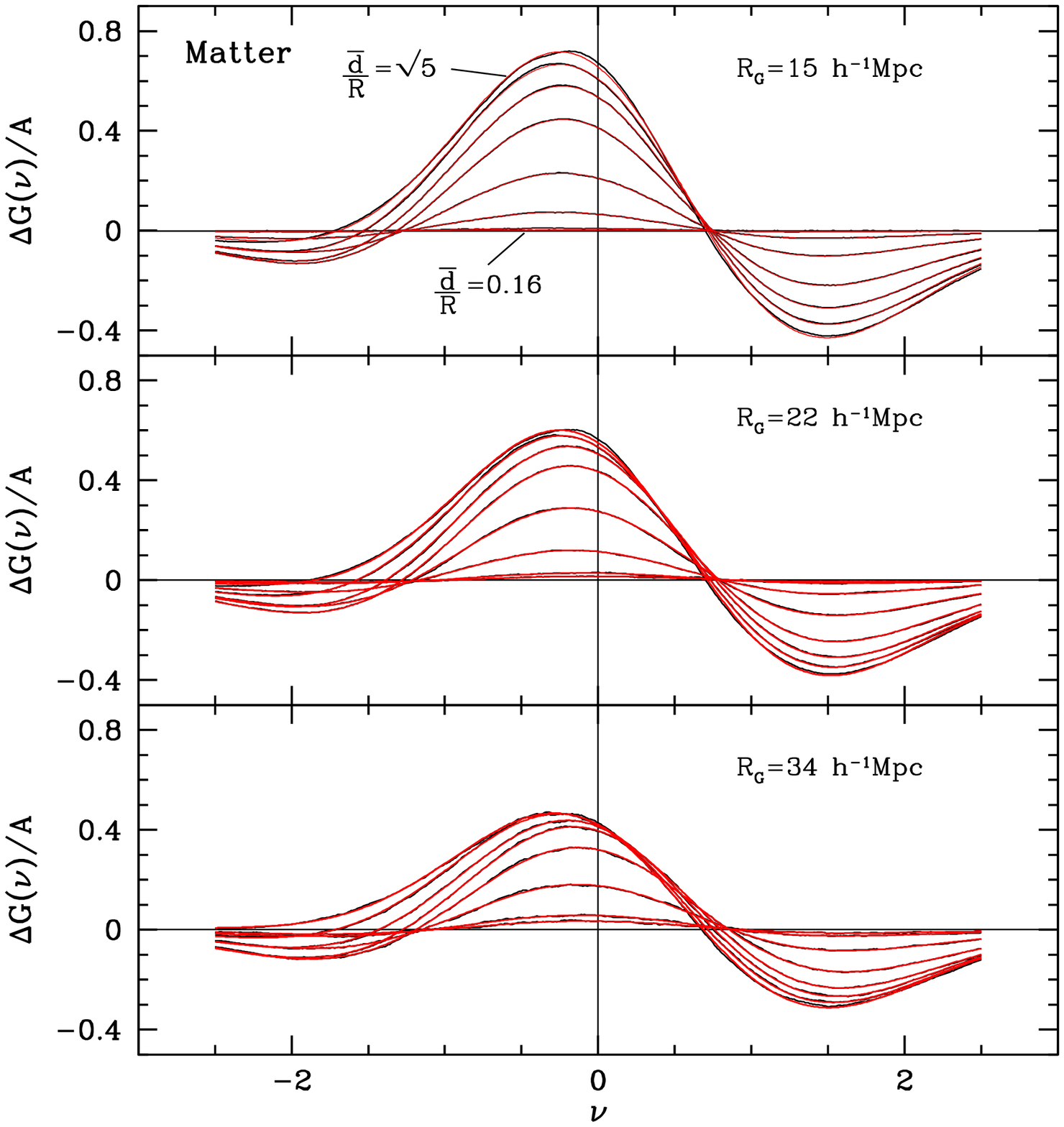}
\caption{
Genus deviation of the matter density fields for 
various mean separations 
($R_G$ = 15, 22, and $34\hMpc$ from the top, middle, and 
bottom panels, respectively). From the top curve in each panel,
$\bar{d}/R_G=\sqrt{5}$, 2,$\sqrt{3}$, $\sqrt{2}$, 1, 0.64, 0.32, and 0.16.
Also we show the fitting curves in the red color.
}
\label{Fig:delgshotm}
\end{figure*}

\begin{figure}
\epsscale{1}
\plotone{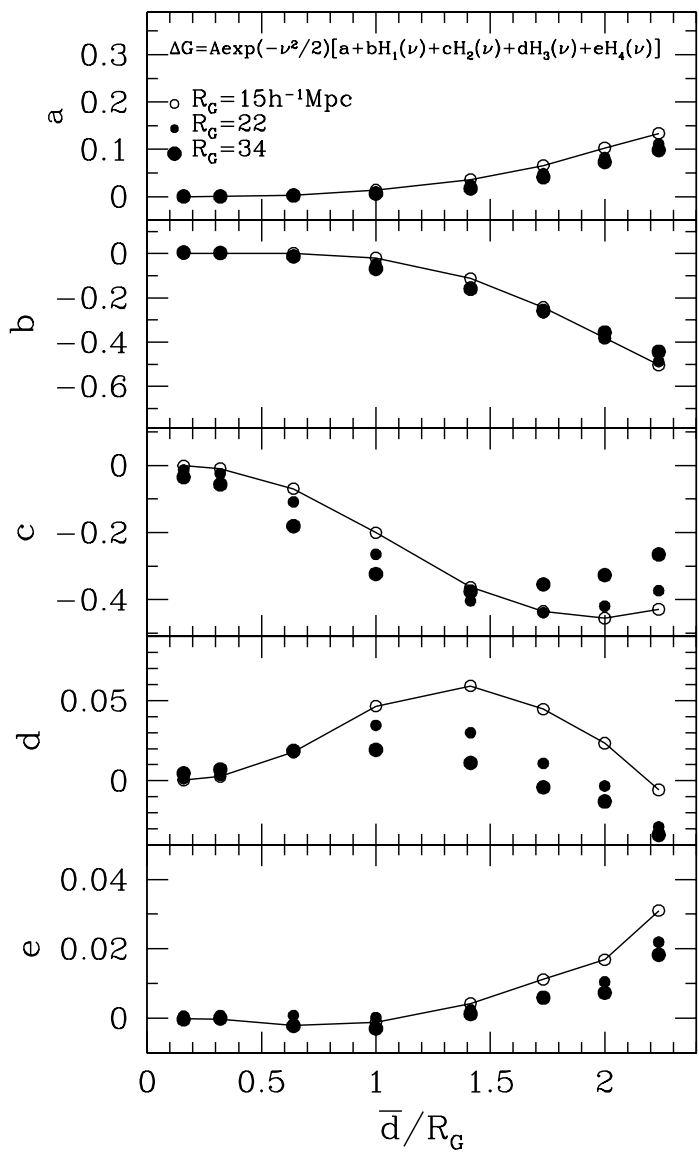}
\caption{
Fitting coefficients $a,b,c,d$ and $e$ for matter density fields.
The open circles are for $R_G$ = 15, small filled circles are for 
$R_G$  = 22, and large filled circles are for $R_G=34\hMpc$. 
}
\label{Fig:fitshotm}
\end{figure}

\begin{figure*}
\epsscale{1}
\plotone{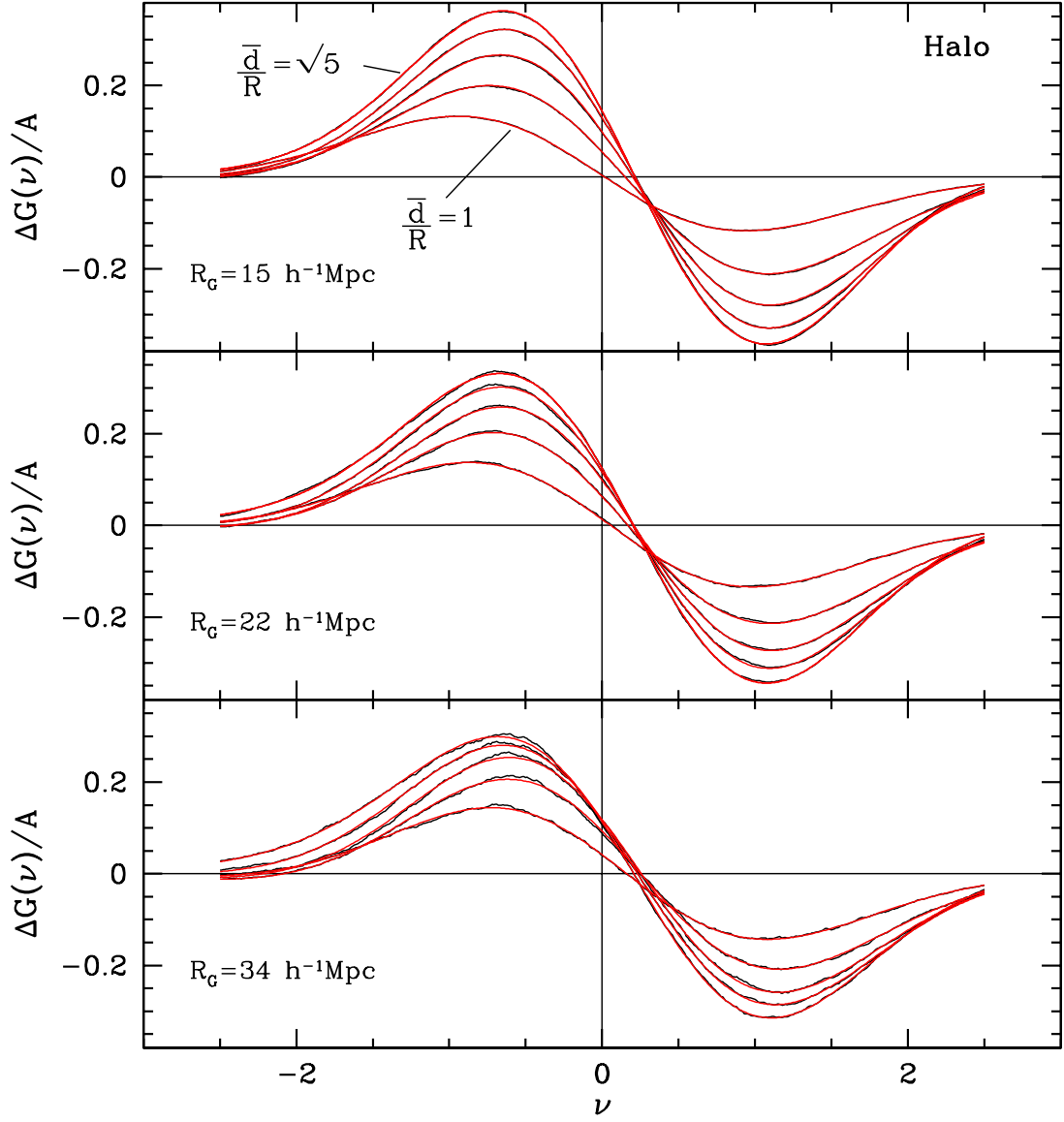}
\caption{
Same as Figure~\ref{Fig:delgshotm}, but for halo density fields.
Y axis is the deviation of halo genus from the 
matter genus of no shot noise. 
From the top curve, $\bar{d}/R=\sqrt{5},2,\sqrt{3},\sqrt{2}$ and 1.}
\label{Fig:delgshoth}
\end{figure*}

\begin{figure}
\epsscale{1}
\plotone{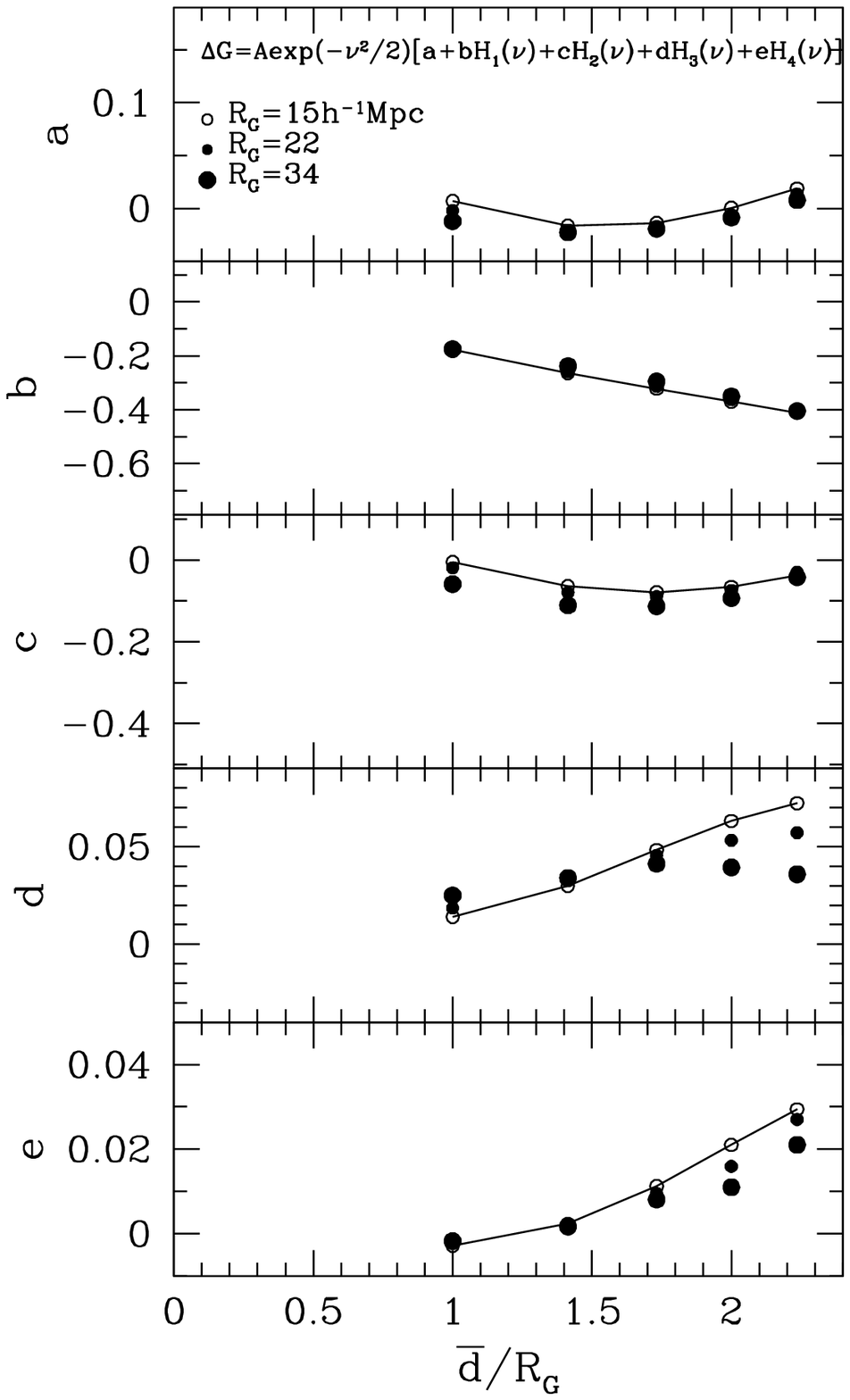}
\caption{
Same as Figure~\ref{Fig:fitshotm}, but for halo density fields.
}
\label{Fig:fitshoth}
\end{figure}

We estimate the genus deviation due to the shot noise 
from $\Delta G_{\rm shot}\equiv G(\nu,\bar{d}/R_G)-G(\nu,\bar{d}/R_G=0)$
with the approximation that
$G(\nu,\bar{d}/R_G=0)\approx G(\nu,\bar{d}/R_G=0.08)$. 
Again, we model this difference using the Hermite polynomials as,
\begin{eqnarray}
\nonumber
&&\Delta G_{\rm shot}(\nu)=A e^{-\nu^2/2}\times \\
&&[aH_0(\nu)+bH_1(\nu)+cH_2(\nu)+dH_3(\nu)+eH_4(\nu)].
\label{Eq:delgshot}
\end{eqnarray}
Figure~\ref{Fig:delgshotm} shows that the model (smooth solid lines in red)
describes the measured genus deviations (black lines) very  well. 
The best-fit coefficients $a, b$ and $e$ are very small for small $\bar{d}/R$  
until $\bar{d}/R \simeq 1.0$ (see Fig.~\ref{Fig:fitshotm}. 
Coefficients $c$ and $d$ move in the opposite directions as
$\bar{d}/R_G$ increases, and the corresponding terms, $cH_2$ and $dH_3$, 
are dominating the genus deviations for $R_G=15h^{-1}$Mpc.

Secondly, we measure the combined effects of shot noise
and halo biasing.
By varying mass cut, we prepare three halo samples having 
the mean separations of $\bar{d}_h$ = 15, 22, and $34\hMpc$. 
The mass cut of each sample in units of $h^{-1}$M$_{\odot}$
is $10^{13.21}$, $10^{13.67}$, and $10^{14.03}$, respectively.

As a result the sparser halo sample has more massive halos that 
are more biased.
We chose to vary the low mass cut to make samples with different 
number density because most observations yield a galaxy redshift sample 
with a faint flux limit or low mass limit.

From each halo sample,
we make four lower-density subsamples by randomly selecting halos so that
the mean halo separation becomes
$\bar{d}/R_G=1$, $\sqrt{2}$, $\sqrt{3}$, 2, and $\sqrt{5}$.
For example, the halo sample with $\bar{d}_h = 15 h^{-1}$Mpc is smoothed
over $R_G=15 h^{-1}$Mpc, and so are its sparser subsamples with $\bar{d}_h =$
$\sqrt{2}, \sqrt{3}$, 2, and $\sqrt{5}$ times $R_G$.
In Table~\ref{Tab:shoth} in Appendix~\ref{App:genus}, the genus values 
are given as a function of $\nu$.

In the bottom panel of Figure~\ref{Fig:gshot}, we show the genus curves of halos for 
$\bar{d}/R_G=1$ (solid) and $\bar{d}/R_G=\sqrt{5}$ (dashed lines). 
The overall amplitude change is smaller than the matter case, but
the shift and asymmetrical change in high and low density regions
show the same pattern.

To measure the combined effects
we calculate the difference of the genus curve between 
the halo and matter density fields 
\begin{equation}
\Delta G_{\rm h,shot}(\nu,\bar{d}/R_G)=G_{\rm h}(\nu,\bar{d}/R_G)
-G_{\rm m}(\nu,\bar{d}/R_G=0),
\end{equation}
where $G_{\rm h}$ and $G_{\rm m}$ are the halo and matter genus curves, respectively.
We use an approximation that  
$G_{\rm m}(\nu,\bar{d}/R_G=0)\approx G_{\rm m}(\nu,\bar{d}/R_G=0.08)$. 
The difference gives us the change of the genus curve
due to the combined effects of the shot noise and halo
bias with respect to the true matter genus curve.
We use Equation~(\ref{Eq:delgshot}) again to fit the deviations. 

In Figure~\ref{Fig:delgshoth} we show the measured genus deviations (black lines) 
as well as the best-fit models (red lines). It can be seen that
Equation~(\ref{Eq:delgshot}) can fit the measurements extremely well again.
Figure~\ref{Fig:fitshoth} shows the variation of the fitting coefficients
as a function of $\bar{d}/R_G$.
Unlike the case of the matter genus curve
the coefficients $a, c, d,$ and $e$ are all rather small, and
the only significant contribution comes from the $bH_1$ term
when $\bar{d}/R_G=1$ (This is the most frequently used choice
for the smoothing length in the study of the genus topology
[\ie Park et al. 2005; Choi et al. 2010]).
The coefficient $b$ is nearly independent of the halo mass cut,
and depends on ${\bar d}/R_G$ linearly.
Due to the term $bH_1$ the genus curve is shifted to the negative
$\nu$'s near the median threshold level.
This is the meat-ball shift for biased objects found by Park \& Gott 
(1991, Fig. 1). The same term also produces an asymmetry of the genus curve
between the high and low density thresholds (Park \& Gott 1991; Park et al. 2005).

Figure~\ref{Fig:fitshoth} suggests that the combined effects of shot noise and halo bias
can be estimated from the data itself because one can measure the slope 
of the linear relation between the coefficient $b$ and $\bar{d}/R_G$ 
using the genus curve obtained for $\bar{d}/R_G=1$ and the condition
that $b=0$ at $\bar{d}/R_G=0$.

%

\section{Summary and Conclusions}

Removing the systematic effects in the genus curve measured from
the observational samples is critically important in the genus topology
analysis. 
Application of the topology analysis in cosmology and galaxy formation 
such as testing non-Gaussianity of the primordial density fluctuations,
constraining the expansion history of the universe, and understanding
the formation and evolution of LSS will be successful 
only after the systematic effects accurately estimated 
from carefully made mock samples are subtracted from the raw data.

Using one of the currently largest cosmological N-body simulations,
the HR2, we accurately measured various systematic effects 
on the genus statistic.
The systematic effects considered are those of finite pixel size,
nonlinear gravitational evolution, redshift-space distortion, 
shot noise or finite sampling, dark matter halo bias.
It is found that all of them can be very accurately modeled with a combination 
of a few low-order Hermite polynomials as follows:
\begin{eqnarray*} 
\Delta G_{\rm pixel}(\nu)&=&A e^{-\nu^2/2}\times\\
&&[aH_0(\nu)+bH_1(\nu)+cH_2(\nu)+dH_4(\nu)]p^2/R_G^2\\
\Delta G_{\rm grav}(\nu)&=&A e^{-\nu^2/2}\times\\
&&[(bH_1(\nu)+dH_3(\nu))\sigma_0+(aH_0(\nu)+cH_2(\nu))\sigma_0^2]\\
\Delta G_{\rm RSD}(\nu)&=&A e^{-\nu^2/2}\times\\
&&[aH_0(\nu)+bH_1(\nu)+cH_2(\nu)+dH_3(\nu)]\\
\Delta G_{\rm h,shot}(\nu)&=&A e^{-\nu^2/2}\times \\
&&[aH_0(\nu)+bH_1(\nu)+cH_2(\nu)+dH_3(\nu)+eH_4(\nu)]
\end{eqnarray*}
where $A$ is the amplitude of the genus curve for a Gaussian field 
with the measured $\sigma_0$ and $\sigma_1$, $p$ is the pixel size, and
$R_G$ is the Gaussian smoothing length.
The first three effects make the genus amplitude decrease, 
but the shot noise, on the other hand, increases it. 
The halo bias mainly makes the genus curve shifted toward the low-density
thresholds without changing the amplitude.

In the pixel effects we find that inclusion of the odd term $H_1$
in addition to $H_0, H_2$, and $H_4$ terms, as suggested by Hamilton et
al. (1986), gives a better fit to the genus deviations.
Figure~\ref{Fig:pixgamp} shows that the pixel size effects should be taken into account
in many topology analyses in order to achieve percent-level accuracy.
Weinberg et al. (1987) has suggested to use a pixel size equal to
or smaller than $p=\lambda /2.5 = \sqrt{2}R_G /2.5 $, and many
follow-up studies adopted this choice.
This choice of pixel size corresponds to the green line in the top
panel of Figure~\ref{Fig:delgpix} when $R_G=15\hMpc$. The figure indicates that
the amplitude drop due to the pixel effect amounts nearly 5\%.
It is desirable to use $p < R_G/3$ and correct the genus curve 
for the remaining pixel effects.

A critical shortcoming of the second-order perturbation theory for
the non-linear gravitational evolution effects is the fact that it does not
predict the amplitude drop of the genus curve, which has long been known
from many numerical studies
(Melott et al. 1988; Park \& Gott 1991; Vegeley et al. 1994; 
Springel et al. 1998; Park et al. 2005; Choi et al. 2010).
We have found that it is the second order term $\sigma_0^2$ in the expansion of 
the genus deviation with the coefficients proportional to $H_0$ and
$H_2$ that are responsible for the amplitude drop of the genus curve
(see Figs.~\ref{Fig:delggrav} and \ref{Fig:fitgrav}).
Therefore, it is necessary to extend the perturbation theory to higher orders
so that the genus deviation can be expanded at least up to the second order
in $\sigma_0$.

The dominant term of the redshift-space distortion effects is the term
proportional to $H_2$ as suggested by Matsubara (1996) from the linear
perturbation theory, and the resulting main systematic effect is 
a decrease in the amplitude of the genus curve. The amount of the decrease
depends on the growth factor and the degree of galaxy bias (Eqs. 9 and 10).
We find that the $H_2$ term is indeed dominant on the smoothing scale of 
$R_G=34\hMpc$ but that the genus deviation should be modeled 
with additional terms such as $H_0, H_1,$ and $H_3$ on smaller scales.
The redshift-space distortion effects make the cluster abundance decrease
more than the void abundance (Figs.~\ref{Fig:gRSD} and \ref{Fig:delgRSD}).
On those scales, it is also expected that the coefficient of $H_2$ is
different from the analytic prediction because the galaxy or halo
bias is not exactly linear and constant as assumed
in the linear perturbation theory (Jee et al. 2012).

The shot noise in the smoothed density field reconstructed from a galaxy catalog
gives the biggest effects on the genus curve and must be very accurately
estimated. 
Figures~\ref{Fig:gshot} and \ref{Fig:delgshotm} 
show that the shot noise characteristically makes 
the amplitude of the genus curve increase by making the density field
choppier, shifts the curve near the median level 
to the left (the meat-ball shift), and increases
the cluster abundance by breaking up high-density regions. 
The void abundance is relatively much less affected.

The shot noise effects diminish greatly when biased objects are used.
The effects become much smaller when the subsample is constructed 
by making the mean halo separation equal to the smoothing length 
through selection of the most biased objects. Figure~\ref{Fig:delgshoth} 
shows that, when ${\bar d}=R_G$,
the genus deviation is symmetric and produces a meat-ball shift in the
genus curve with little change in amplitude.
Considering the characteristics of the shot noise effects, we conclude that
this deviation proportional to $H_1$ (Fig.~\ref{Fig:fitshoth}) 
is mostly due to the halo bias.
The figure indicates that the optimal Gaussian smoothing length 
for the genus topology analysis is $R_G = {\bar d}$, 
the mean halo/galaxy separation itself. 
In the early studies of topology of large-scale structure 
the choice of $R_G = {\bar d}/\sqrt{2}$ has been widely used 
(Weinberg et al. 1987; Gott et al. 1989; Vogeley et al. 1994).
But $R_G = {\bar d}$ is used in most recent analyses (Park et al. 2005;
Gott et al. 2009; Choi et al. 2010, 2014).

We suggest not to use subsamples with galaxies randomly selected
from the parent catalog as this greatly amplifies the shot noise effects.
Instead, it is desirable to vary the luminosity or mass cut
to make subsamples of galaxies.
The sample used in the top panel of Figure~\ref{Fig:delgshoth} contains halos 
with a low mass cut that results in the mean halo separation of $15\hMpc$ 
while the halo sample in the bottom panel has a relatively higher low mass cut
giving ${\bar d}=34\hMpc$. If we randomly remove some halos in the first
sample to make the mean separation $\sqrt{5}\times 15\hMpc\approx 34\hMpc$, 
the amplitude excess due to the shot noise is about 15\% at $\nu=0$. 
On the other hand, it is about 4\% for the latter sample of more massive halos.

It should be pointed out that our estimation of the systematic effects
is directly useful only for the particular $\Lambda$CDM model 
we adopt and for a set of particular smoothing scales and dark halo mass cuts.
When one analyzes an observational sample and compare it with
a cosmological model, it is necessary to perform a detailed modeling
of the observed objects in that particular cosmology and estimate
all the systematic effects in the genus curve as considered in this paper.
Similar studies are needed for other Minkowski functionals and Betti 
numbers.

\acknowledgments
The authors thank 
Prof. Juhan Kim for helpful comments on the paper
and also for kindly providing a parallel code for computing the genus.
The authors also thank Korea Institute for
for Advanced Study for providing computing resources
(KIAS Center for Advanced Computation Linux Cluster
System).
Y.-Y. C was supported
by the National Research Foundation of Korea to the Center
for Galaxy Evolution Research. (NO. 2010-0027910). J.-E.L.
was supported by the 2013 Sabbatical Leave Program of Kyung
Hee University (KHU-20131724).
S.S.K., J.S., and M.K.'s work was supported by the BK21 plus program through the
NRF of Korea.

\appendix
\section{Genus Curves}\label{App:genus}
In this appendix, we provide tables of the genus curves used in this study
for the reader who wishes to study the various systematic effects on the genus curve.
All the systematic effects were examined at 
three different smoothing lengths, $R_G$ = 15, 22, and $34\hMpc$.
Electronic versions of these tables are
available from the authors upon request.
\begin{deluxetable}{rrrrrrrrrrrrr}
\tabletypesize{\tiny}
\tablecaption{
Genus Values at a Given Threshold Level of
the Matter Density Field Sample Calculated for Various
Pixel Sizes \label{Tab:pix}}
\tablewidth{0pt}
\tablehead{
\colhead{$\nu$} &\colhead{$p=3.33$} & \colhead{3.75} & \colhead{4 }& 
\colhead{5}     &\colhead{5.71}     & \colhead{6}    & \colhead{6.67} & 
\colhead{7.5}   &\colhead{8}        & \colhead{8.57} & \colhead{10} & 
\colhead{12}
}
\startdata
\cutinhead{$R_G=15\hMpc$}
-2.5 & -108992 & -108446 &  -108034 & -106431 & -105145 & -104239 & -102867 & -101024 & -99655 & -97919 & -93888 &-87611 \\
-2.4 & -126596 & -126014 &  -125672 & -123654 & -122118 & -121500 & -119731 & -117498 & -116077 & -114136 & -110032 &-102595 \\
-2.3 & -144775 & -144054 &  -143714 & -141749 & -140030 & -139400 & -137685 & -135152 & -133547 & -131517 & -126524 &-118766 \\
-2.2 & -162787 & -162182 &  -161581 & -159410 & -157800 & -156919 & -155079 & -152367 & -150773 & -148444 & -143278 &-134976 \\
-2.1 & -179620 & -178842 &  -178345 & -176289 & -174534 & -173733 & -171781 & -168850 & -167352 & -165167 & -159289 &-150682 \\
-2.0 & -193867 & -193025 &  -192666 & -190433 & -188558 & -187802 & -185993 & -183191 & -181408 & -179469 & -173579 &-164596 \\
-1.9 & -205228 & -204550 &  -204083 & -201908 & -199996 & -199355 & -197337 & -194668 & -192764 & -190457 & -184843 &-176037 \\
-1.8 & -212676 & -211876 &  -211475 & -209243 & -207460 & -206677 & -204859 & -202168 & -200383 & -198058 & -192746 &-183824 \\
-1.7 & -214571 & -213864 &  -213361 & -211382 & -209784 & -208815 & -207271 & -204568 & -202750 & -201042 & -195581 &-187241 \\
-1.6 & -209091 & -208573 &  -208015 & -206367 & -204784 & -204072 & -202377 & -200072 & -198606 & -196872 & -192190 &-184529 \\
-1.5 & -195988 & -195411 &  -194965 & -193413 & -192057 & -191581 & -190031 & -188248 & -186937 & -185502 & -181706 &-174827 \\
-1.4 & -174461 & -174139 &  -173879 & -172614 & -171296 & -171135 & -170058 & -168425 & -167225 & -166297 & -162801 &-157825 \\
-1.3 & -143570 & -143278 &  -143142 & -142274 & -141460 & -141565 & -140721 & -139763 & -139031 & -138275 & -135869 &-132505 \\
-1.2 & -103593 & -103612 &  -103408 & -102974 & -102802 & -102905 & -102200 & -101933 & -101296 & -101049 & -100399 &-98802 \\
-1.1 &  -56058 &  -56061 &  -56270 & -56375 & -56188 & -56428 & -56260 & -56186 & -56243 & -56132 & -56523 &-57334 \\
-1.0 &    -407 &    -612 &  -657 & -1527 & -2056 & -2050 & -2934 & -3940 & -4308 & -4818 & -6365 &-8967 \\
-0.9 &   60985 &   60583 &  60326 & 59440 & 58763 & 58290 & 56959 & 55766 & 54705 & 54115 & 50553 &46207 \\
-0.8 &  126949 &  126572 &  126266 & 124489 & 123460 & 122904 & 121273 & 119249 & 117717 & 115904 & 111555 &104906 \\
-0.7 &  194587 &  193832 &  193356 & 191563 & 189561 & 188995 & 187096 & 184283 & 182836 & 180260 & 175161 &165835 \\
-0.6 &  260604 &  259723 &  259473 & 256906 & 254749 & 254096 & 251310 & 248538 & 246509 & 243888 & 237214 &226104 \\
-0.5 &  323216 &  322339 &  321560 & 318851 & 316517 & 315362 & 312734 & 309152 & 306863 & 303586 & 295844 &282960 \\
-0.4 &  378505 &  377658 &  376817 & 373660 & 371139 & 369905 & 367541 & 363531 & 360868 & 357536 & 348597 &334381 \\
-0.3 &  425471 &  424269 &  423642 & 420387 & 417370 & 416336 & 413033 & 408785 & 405814 & 402505 & 393010 &377732 \\
-0.2 &  460367 &  459416 &  458641 & 455393 & 452324 & 451290 & 447863 & 443601 & 440644 & 437105 & 426838 &411383 \\
-0.1 &  483040 &  481781 &  481127 & 477782 & 474795 & 473449 & 470087 & 465719 & 462712 & 458947 & 448575 &432032 \\
0.0 &   490296 &  489250 &  488562 & 485161 & 482177 & 481141 & 477891 & 473071 & 470221 & 466574 & 456580 &439839 \\
0.1 &   482916 &  481940 &  481321 & 477947 & 474981 & 473839 & 470434 & 465854 & 463011 & 459939 & 449629 &433701 \\
0.2 &   461376 &  460432 &  459785 & 456635 & 453875 & 452460 & 449318 & 445536 & 442702 & 438964 & 429614 &414991 \\
0.3 &   426492 &  425592 &  425226 & 421808 & 419721 & 418175 & 415789 & 411481 & 409055 & 406090 & 397301 &383053 \\
0.4 &   379352 &  378466 &  378044 & 375348 & 372996 & 372046 & 369857 & 365953 & 363951 & 361020 & 353085 &340592 \\
0.5 &   323075 &  322211 &  321747 & 319275 & 317760 & 316645 & 314426 & 311411 & 309628 & 307084 & 300007 &289172 \\
0.6 &   260303 &  259656 &  259411 & 257194 & 255706 & 255257 & 253403 & 250919 & 249484 & 247764 & 241514 &232562 \\
0.7 &   192725 &  192475 &  192188 & 190761 & 189602 & 189276 & 187810 & 185887 & 184930 & 183364 & 178411 &171313 \\
0.8 &   124309 &  124108 &  123981 & 122858 & 122439 & 122008 & 120961 & 119699 & 119209 & 118272 & 114651 &109352 \\
0.9 &    57185 &   56921 &  56746 & 56345 & 55915 & 55935 & 55158 & 54152 & 53684 & 53249 & 51825 &48740 \\
1.0 &    -6323 &   -6404 &  -6434 & -6378 & -6470 & -6210 & -5944 & -6452 & -6362 & -6146 & -7049 &-7305 \\
1.1 &   -63409 &  -63117 &  -62873 & -62538 & -62466 & -61959 & -61348 & -61090 & -60797 & -60512 & -59350 &-57861 \\
1.2 &  -113185 & -112725 &  -112436 & -111564 & -110637 & -110295 & -109507 & -108490 & -107585 & -106832 & -104549 &-101385 \\
1.3 &  -153819 & -153339 &  -153066 & -151912 & -150991 & -150218 & -149249 & -147383 & -146112 & -145156 & -141725 &-137111 \\
1.4 &  -186058 & -185560 &  -185164 & -183600 & -181853 & -181486 & -179859 & -177693 & -176376 & -174824 & -170356 &-163804 \\
1.5 &  -208250 & -207671 &  -207251 & -205215 & -203526 & -202641 & -201024 & -198601 & -196929 & -195062 & -190242 &-182128 \\
1.6 &  -221960 & -221228 &  -220801 & -218443 & -216553 & -215916 & -213630 & -210975 & -209410 & -207174 & -201648 &-193002 \\
1.7 &  -227736 & -226832 &  -226407 & -223830 & -221988 & -221003 & -218978 & -215789 & -213927 & -211672 & -205532 &-195643 \\
1.8 &  -226079 & -225257 &  -224548 & -222028 & -219895 & -218926 & -216719 & -213639 & -211642 & -209250 & -203001 &-193162 \\
1.9 &  -218582 & -217717 &  -217275 & -214663 & -212539 & -211626 & -209280 & -206183 & -204055 & -201854 & -195114 &-185223 \\
2.0 &  -206241 & -205450 &  -204737 & -202372 & -200310 & -199285 & -196867 & -194012 & -192022 & -189732 & -183237 &-173233 \\
2.1 &  -190950 & -190159 &  -189609 & -187252 & -185299 & -184255 & -182012 & -179171 & -177117 & -174734 & -168586 &-158923 \\
2.2 &  -173112 & -172436 &  -171849 & -169653 & -167638 & -166709 & -164590 & -161988 & -159853 & -157671 & -151761 &-142646 \\
2.3 &  -153979 & -153169 &  -152791 & -150634 & -148799 & -148030 & -145971 & -143552 & -141661 & -139672 & -134318 &-125924 \\
2.4 &  -134751 & -134052 &  -133634 & -131588 & -129943 & -129311 & -127540 & -125169 & -123608 & -121753 & -116816 &-109248 \\
2.5 &  -116288 & -115712 &  -115387 & -113582 & -111788 & -111586 & -109909 & -107631 & -106360 & -104843 & -100168 &-93502 \\
\cutinhead{$R_G=22\hMpc$}
-2.5 & -41134 &-41018 & -40963 & -40534 & -40404 & -40281 & -40154 & -39598 & -39395 & -39059 & -38185 &-36865 \\
-2.4 & -47665 &-47533 & -47412 & -47081 & -46899 & -46686 & -46484 & -45955 & -45649 & -45388 & -44449 &-43010 \\
-2.3 & -54593 &-54482 & -54384 & -53988 & -53756 & -53599 & -53386 & -52776 & -52605 & -52151 & -51137 &-49521 \\
-2.2 & -61261 &-61105 & -61109 & -60681 & -60424 & -60209 & -59917 & -59426 & -59104 & -58718 & -57686 &-56084 \\
-2.1 & -67562 &-67443 & -67360 & -66959 & -66668 & -66565 & -66229 & -65636 & -65375 & -64922 & -63800 &-61995 \\
-2.0 & -73024 &-72907 & -72804 & -72391 & -72053 & -71909 & -71546 & -71100 & -70683 & -70295 & -69267 &-67452 \\
-1.9 & -77381 &-77197 & -77123 & -76724 & -76395 & -76176 & -75833 & -75293 & -75013 & -74603 & -73539 &-71752 \\
-1.8 & -79818 &-79691 & -79640 & -79269 & -78842 & -78830 & -78407 & -78045 & -77620 & -77246 & -76202 &-74416 \\
-1.7 & -80154 &-80101 & -80057 & -79633 & -79295 & -79303 & -78876 & -78334 & -78174 & -77794 & -76671 &-75117 \\
-1.6 & -77793 &-77731 & -77660 & -77365 & -77097 & -76937 & -76661 & -76264 & -75970 & -75707 & -74796 &-73519 \\
-1.5 & -72758 &-72617 & -72563 & -72266 & -72060 & -72032 & -71810 & -71406 & -71216 & -71031 & -70047 &-69023 \\
-1.4 & -64189 &-64196 & -64172 & -63943 & -63876 & -63753 & -63607 & -63340 & -63221 & -62958 & -62439 &-61352 \\
-1.3 & -52371 &-52335 & -52340 & -52220 & -52152 & -52209 & -52020 & -51866 & -51668 & -51470 & -51106 &-50631 \\
-1.2 & -37173 &-37159 & -37070 & -37031 & -36991 & -37026 & -37146 & -36938 & -37000 & -36865 & -36675 &-36526 \\
-1.1 & -18771 &-18749 & -18730 & -18824 & -18813 & -18878 & -18885 & -18961 & -18960 & -18902 & -19003 &-19282 \\
-1.0 &   2204 &  2208 & 2173 & 2074 & 1942 & 1885 & 1794 & 1570 & 1379 & 1139 & 953 &289 \\
-0.9 &  25406 & 25328 & 25284 & 25177 & 25073 & 24915 & 24775 & 24548 & 24313 & 24153 & 23582 &22601 \\
-0.8 &  49929 & 49839 & 49847 & 49485 & 49234 & 49163 & 48908 & 48493 & 48169 & 47975 & 47115 &45746 \\
-0.7 &  75260 & 75146 & 75066 & 74624 & 74281 & 74172 & 73796 & 73367 & 73018 & 72772 & 71604 &70031 \\
-0.6 & 100192 &100030 & 99987 & 99524 & 99227 & 99096 & 98659 & 98054 & 97636 & 97071 & 95949 &94122 \\
-0.5 & 123912 &123767 & 123651 & 123154 & 122814 & 122529 & 122070 & 121310 & 121110 & 120494 & 118955 &116389 \\
-0.4 & 145023 &144812 & 144733 & 144115 & 143660 & 143435 & 142861 & 142166 & 141751 & 141116 & 139624 &136978 \\
-0.3 & 161794 &161612 & 161472 & 160928 & 160339 & 160168 & 159619 & 158898 & 158337 & 157864 & 156000 &153285 \\
-0.2 & 174460 &174217 & 174054 & 173555 & 172968 & 172847 & 172255 & 171366 & 170779 & 170133 & 168692 &165819 \\
-0.1 & 181982 &181705 & 181661 & 180963 & 180481 & 180264 & 179676 & 178830 & 178335 & 177936 & 175927 &173193 \\
0.0 &  184278 &184128 & 184058 & 183450 & 182948 & 182613 & 182101 & 181341 & 180770 & 180137 & 178249 &175716 \\
0.1 &  181546 &181294 & 181192 & 180622 & 180169 & 179984 & 179356 & 178689 & 178006 & 177522 & 175790 &172913 \\
0.2 &  173054 &172883 & 172786 & 172281 & 171843 & 171506 & 171066 & 170329 & 169785 & 169295 & 167601 &164929 \\
0.3 &  159963 &159809 & 159660 & 159123 & 158679 & 158401 & 157976 & 157181 & 156669 & 156047 & 154574 &151864 \\
0.4 &  142114 &141989 & 141773 & 141449 & 140923 & 140808 & 140249 & 139600 & 139188 & 138714 & 137282 &134864 \\
0.5 &  120685 &120481 & 120426 & 119995 & 119733 & 119538 & 119128 & 118508 & 118092 & 117696 & 116523 &114276 \\
0.6 &   96707 & 96524 & 96500 & 96166 & 95875 & 95707 & 95462 & 94740 & 94702 & 94478 & 93168 &91474 \\
0.7 &   71154 & 71151 & 71122 & 70856 & 70514 & 70536 & 70157 & 69872 & 69741 & 69244 & 68505 &67425 \\
0.8 &   44985 & 44965 & 44929 & 44749 & 44482 & 44528 & 44411 & 44156 & 44044 & 43806 & 43201 &42194 \\
0.9 &   19694 & 19665 & 19681 & 19594 & 19583 & 19539 & 19438 & 19368 & 19261 & 19056 & 18797 &18012 \\
1.0 &   -4038 & -4091 & -4091 & -4091 & -4098 & -4073 & -4319 & -4106 & -4196 & -4388 & -4429 &-4571 \\
1.1 &  -25481 &-25488 & -25481 & -25357 & -25291 & -25317 & -25374 & -25064 & -25145 & -25117 & -24814 &-24452 \\
1.2 &  -43670 &-43612 & -43590 & -43402 & -43295 & -43238 & -43216 & -42889 & -42791 & -42676 & -42260 &-41623 \\
1.3 &  -59174 &-59082 & -59016 & -58808 & -58639 & -58470 & -58346 & -58010 & -57907 & -57440 & -56818 &-55855 \\
1.4 &  -70850 &-70808 & -70643 & -70459 & -70086 & -69985 & -69729 & -69392 & -69116 & -68765 & -67976 &-66628 \\
1.5 &  -78835 &-78741 & -78688 & -78332 & -78068 & -77951 & -77626 & -77202 & -76929 & -76561 & -75633 &-74110 \\
1.6 &  -83970 &-83840 & -83763 & -83355 & -83056 & -82922 & -82586 & -82053 & -81714 & -81329 & -80195 &-78224 \\
1.7 &  -85950 &-85854 & -85771 & -85403 & -85001 & -84877 & -84469 & -83802 & -83481 & -83085 & -81853 &-79851 \\
1.8 &  -85257 &-85100 & -84916 & -84522 & -84031 & -83932 & -83500 & -83014 & -82500 & -82043 & -80781 &-78838 \\
1.9 &  -82259 &-82127 & -81952 & -81527 & -81033 & -80941 & -80585 & -79952 & -79548 & -79125 & -77885 &-75789 \\
2.0 &  -77367 &-77233 & -77059 & -76686 & -76245 & -76094 & -75661 & -74975 & -74738 & -74271 & -73048 &-71042 \\
2.1 &  -71645 &-71466 & -71405 & -70968 & -70573 & -70402 & -69989 & -69508 & -69102 & -68625 & -67439 &-65367 \\
2.2 &  -64801 &-64680 & -64594 & -64159 & -63878 & -63651 & -63268 & -62698 & -62404 & -61892 & -60656 &-58838 \\
2.3 &  -57654 &-57518 & -57437 & -57079 & -56784 & -56612 & -56156 & -55743 & -55372 & -55092 & -53916 &-52285 \\
2.4 &  -50693 &-50593 & -50510 & -50147 & -49847 & -49671 & -49309 & -48943 & -48617 & -48293 & -47224 &-45751 \\
2.5 &  -43807 &-43704 & -43652 & -43326 & -43067 & -42918 & -42646 & -42195 & -41930 & -41587 & -40620 &-39232 \\
\cutinhead{$R_G=34\hMpc$}
-2.5 & -13250 & -13225&  -13226 & -13182 & -13179 & -13153 & -13108 & -13059 & -13009 & -12945 & -12818 &-12622 \\
-2.4 & -15377 & -15357&  -15371 & -15318 & -15267 & -15270 & -15216 & -15163 & -15133 & -15077 & -14957 &-14757 \\
-2.3 & -17462 & -17445&  -17430 & -17365 & -17327 & -17329 & -17290 & -17222 & -17183 & -17135 & -17014 &-16792 \\
-2.2 & -19491 & -19473&  -19477 & -19417 & -19350 & -19348 & -19314 & -19234 & -19213 & -19163 & -18973 &-18743 \\
-2.1 & -21464 & -21457&  -21430 & -21410 & -21346 & -21328 & -21305 & -21237 & -21189 & -21122 & -21004 &-20766 \\
-2.0 & -23147 & -23116&  -23101 & -23020 & -22990 & -22952 & -22925 & -22846 & -22800 & -22739 & -22556 &-22306 \\
-1.9 & -24485 & -24479&  -24472 & -24395 & -24332 & -24323 & -24278 & -24219 & -24158 & -24100 & -23963 &-23722 \\
-1.8 & -25212 & -25201&  -25191 & -25118 & -25089 & -25058 & -25002 & -24936 & -24837 & -24814 & -24660 &-24438 \\
-1.7 & -25175 & -25171&  -25158 & -25111 & -25050 & -25038 & -25006 & -24921 & -24897 & -24870 & -24738 &-24455 \\
-1.6 & -24390 & -24383&  -24362 & -24357 & -24347 & -24325 & -24303 & -24233 & -24179 & -24132 & -24038 &-23903 \\
-1.5 & -22789 & -22786&  -22789 & -22740 & -22685 & -22704 & -22636 & -22595 & -22628 & -22546 & -22498 &-22308 \\
-1.4 & -20156 & -20155&  -20147 & -20128 & -20107 & -20144 & -20115 & -20083 & -20035 & -20042 & -19932 &-19801 \\
-1.3 & -16294 & -16311&  -16276 & -16268 & -16287 & -16269 & -16210 & -16193 & -16232 & -16239 & -16203 &-16177 \\
-1.2 & -11414 & -11407&  -11421 & -11399 & -11405 & -11426 & -11396 & -11411 & -11386 & -11322 & -11355 &-11271 \\
-1.1 &  -5758 &  -5753&  -5803 & -5761 & -5730 & -5753 & -5734 & -5720 & -5770 & -5744 & -5871 &-5850 \\
-1.0 &    890 &    877&  862 & 847 & 865 & 898 & 833 & 801 & 757 & 761 & 715 &717 \\
-0.9 &   8114 &   8122&  8143 & 8114 & 8093 & 8121 & 8070 & 8039 & 8001 & 8022 & 7892 &7767 \\ 
-0.8 &  15838 &  15828&  15819 & 15744 & 15762 & 15752 & 15757 & 15642 & 15620 & 15704 & 15534 &15343 \\
-0.7 &  23770 &  23789&  23749 & 23692 & 23675 & 23631 & 23562 & 23503 & 23497 & 23453 & 23260 &23066 \\
-0.6 &  31735 &  31711&  31694 & 31643 & 31593 & 31584 & 31517 & 31446 & 31444 & 31355 & 31178 &30902 \\
-0.5 &  39136 &  39116&  39117 & 39030 & 38992 & 38945 & 38896 & 38893 & 38792 & 38775 & 38623 &38302 \\
-0.4 &  45995 &  45947&  45920 & 45848 & 45790 & 45733 & 45710 & 45608 & 45586 & 45525 & 45331 &45068 \\
-0.3 &  51434 &  51410&  51392 & 51328 & 51258 & 51235 & 51161 & 51029 & 51026 & 50912 & 50737 &50289 \\
-0.2 &  55378 &  55370&  55368 & 55244 & 55215 & 55175 & 55093 & 55012 & 54945 & 54848 & 54653 &54304 \\
-0.1 &  57771 &  57734&  57709 & 57638 & 57560 & 57524 & 57439 & 57354 & 57279 & 57147 & 56962 &56532 \\
0.0 &   58438 &  58435&  58403 & 58344 & 58263 & 58269 & 58172 & 58089 & 58059 & 57948 & 57802 &57383 \\
0.1 &   57578 &  57538&  57497 & 57407 & 57369 & 57329 & 57273 & 57146 & 57110 & 56973 & 56803 &56494 \\
0.2 &   54826 &  54796&  54799 & 54714 & 54650 & 54689 & 54555 & 54518 & 54400 & 54340 & 54108 &53789 \\
0.3 &   50461 &  50429&  50446 & 50353 & 50315 & 50285 & 50220 & 50125 & 50046 & 49940 & 49775 &49475 \\
0.4 &   44616 &  44592&  44591 & 44533 & 44445 & 44423 & 44371 & 44308 & 44241 & 44157 & 43942 &43652 \\
0.5 &   37705 &  37691&  37721 & 37608 & 37530 & 37515 & 37510 & 37388 & 37370 & 37332 & 37144 &36763 \\
0.6 &   30136 &  30124&  30113 & 30089 & 30042 & 30032 & 29973 & 29892 & 29911 & 29881 & 29740 &29537 \\
0.7 &   22292 &  22289&  22274 & 22242 & 22187 & 22176 & 22107 & 22115 & 22043 & 22029 & 21898 &21616 \\
0.8 &   14091 &  14103&  14103 & 14021 & 14043 & 13995 & 13994 & 13984 & 13947 & 13938 & 13826 &13610 \\
0.9 &    6205 &   6193&  6224 & 6166 & 6117 & 6199 & 6148 & 6136 & 6146 & 6106 & 6081 &5979 \\ 
1.0 &   -1353 &  -1354&  -1358 & -1323 & -1399 & -1347 & -1442 & -1425 & -1433 & -1428 & -1493 &-1455 \\
1.1 &   -7988 &  -7997&  -7997 & -7997 & -8006 & -7973 & -8026 & -7987 & -8013 & -7984 & -7942 &-7997 \\
1.2 &  -13831 & -13815&  -13801 & -13777 & -13819 & -13777 & -13748 & -13762 & -13744 & -13726 & -13773 &-13738 \\
1.3 &  -18649 & -18639&  -18629 & -18607 & -18600 & -18574 & -18543 & -18517 & -18545 & -18424 & -18390 &-18267 \\
1.4 &  -22263 & -22266&  -22237 & -22216 & -22189 & -22152 & -22111 & -22030 & -22010 & -22021 & -21826 &-21672 \\
1.5 &  -24713 & -24707&  -24698 & -24631 & -24635 & -24607 & -24563 & -24501 & -24495 & -24430 & -24312 &-24115 \\
1.6 &  -26366 & -26339&  -26323 & -26306 & -26277 & -26233 & -26180 & -26145 & -26125 & -26058 & -25907 &-25612 \\
1.7 &  -26923 & -26892&  -26888 & -26824 & -26800 & -26773 & -26721 & -26637 & -26629 & -26525 & -26395 &-26180 \\
1.8 &  -26776 & -26754&  -26736 & -26658 & -26598 & -26588 & -26512 & -26465 & -26373 & -26314 & -26152 &-25906 \\
1.9 &  -25838 & -25817&  -25804 & -25725 & -25670 & -25673 & -25596 & -25525 & -25441 & -25428 & -25229 &-24929 \\
2.0 &  -24369 & -24352&  -24343 & -24259 & -24207 & -24197 & -24117 & -24034 & -23981 & -23918 & -23721 &-23380 \\
2.1 &  -22425 & -22384&  -22379 & -22310 & -22231 & -22208 & -22157 & -22072 & -22009 & -21941 & -21814 &-21523 \\
2.2 &  -20325 & -20308&  -20281 & -20229 & -20193 & -20133 & -20116 & -20032 & -19967 & -19906 & -19747 &-19471 \\
2.3 &  -18180 & -18167&  -18130 & -18096 & -18072 & -18012 & -18006 & -17908 & -17849 & -17791 & -17643 &-17426 \\
2.4 &  -16031 & -16015&  -15996 & -15946 & -15905 & -15866 & -15856 & -15759 & -15721 & -15644 & -15497 &-15287 \\
2.5 &  -13815 & -13805&  -13806 & -13749 & -13696 & -13699 & -13626 & -13585 & -13544 & -13498 & -13317 &-13123 \\
\enddata
\tablecomments{
Genus values at a given threshold level of the density field estimated from a discrete particle distribution. The density field was smoothed with smoothing length $R_G=15\hMpc$.
We use a set of grids with cubic pixels and with size of 2160, 1920, 1800, 1440, 1260, 1200, 1080, 960, 900, 840, 720, and 600
in terms of number of pixels along a side.Since the physical size of HR2 is 7200 $\hMpc$, they correspond to
3.33, 3.75, 4, 5, 5.71, 6, 6.67, 7.5, 8, 8.57, 10, and $12\hMpc$,respectively, in terms of pixel size $p$.}
\end{deluxetable}

\begin{table*}
\caption{
Genus values at a Given Threshold Level for the Matter Density Field Samples
Used to Estimate the Non-linear
Gravitational Evolution Effects
}\label{Tab:grav}
\tiny
\centering
\begin{center}
\begin{tabular}{r|rr|rr|rr}
\hline\hline
        &\multicolumn{2}{c|}{$R_G=15h^{-1}$Mpc} &\multicolumn{2}{c|}{$R_G=22h^{-1}$Mpc} &\multicolumn{2}{c}{$R_G=34h^{-1}$Mpc}\\
$\nu$ & $G_{\rm m,r}|_{z=32}$ & $G_{\rm m,r}|_{z=0}$
      & $G_{\rm m,r}|_{z=32}$ & $G_{\rm m,r}|_{z=0}$
      & $G_{\rm m,r}|_{z=32}$ & $G_{\rm m,r}|_{z=0}$ \\
\hline
-2.5 & -108992  & -116313  & -41134  & -43267 &-13250  &-13683\\
-2.4 & -126596  & -135266  & -47665  & -50342 &-15377  &-15901\\
-2.3 & -144775  & -154571  & -54593  & -57606 &-17462  &-18078\\
-2.2 & -162787  & -173653  & -61261  & -64555 &-19491  &-20234\\
-2.1 & -179620  & -190606  & -67562  & -71101 &-21464  &-22285\\
-2.0 & -193867  & -206042  & -73024  & -76917 &-23147  &-24115\\
-1.9 & -205228  & -218159  & -77381  & -81628 &-24485  &-25448\\
-1.8 & -212676  & -225541  & -79818  & -84236 &-25212  &-26240\\
-1.7 & -214571  & -226813  & -80154  & -84723 &-25175  &-26353\\
-1.6 & -209091  & -220701  & -77793  & -82486 &-24390  &-25706\\
-1.5 & -195988  & -207088  & -72758  & -77269 &-22789  &-23980\\
-1.4 & -174461  & -184193  & -64189  & -68525 &-20156  &-21381\\
-1.3 & -143570  & -151186  & -52371  & -56390 &-16294  &-17438\\
-1.2 & -103593  & -109502  & -37173  & -40876 &-11414  &-12575\\
-1.1 &  -56058  &  -58746  & -18771  & -21605 & -5758  & -6745\\
-1.0 &    -407  &    -307  &   2204  &   -104 &   890  &   -68\\
-0.9 &   60985  &   64655  &  25406  &  23958 &  8114  &  7232\\
-0.8 &  126949  &  133567  &  49929  &  49094 & 15838  & 15138\\
-0.7 &  194587  &  203207  &  75260  &  75396 & 23770  & 23416\\
-0.6 &  260604  &  272365  & 100192  & 101707 & 31735  & 31371\\
-0.5 &  323216  &  337302  & 123912  & 125978 & 39136  & 39205\\
-0.4 &  378505  &  395286  & 145023  & 147247 & 45995  & 46214\\
-0.3 &  425471  &  444378  & 161794  & 164908 & 51434  & 51848\\
-0.2 &  460367  &  480671  & 174460  & 178298 & 55378  & 56050\\
-0.1 &  483040  &  503357  & 181982  & 186729 & 57771  & 58542\\
 0.0 &  490296  &  510839  & 184278  & 190051 & 58438  & 59482\\
 0.1 &  482916  &  503491  & 181546  & 186997 & 57578  & 58662\\
 0.2 &  461376  &  480720  & 173054  & 178788 & 54826  & 56074\\
 0.3 &  426492  &  444352  & 159963  & 165242 & 50461  & 51751\\
 0.4 &  379352  &  395592  & 142114  & 147385 & 44616  & 45881\\
 0.5 &  323075  &  337400  & 120685  & 125919 & 37705  & 38921\\
 0.6 &  260303  &  272285  &  96707  & 101880 & 30136  & 31760\\
 0.7 &  192725  &  203385  &  71154  &  75822 & 22292  & 23580\\
 0.8 &  124309  &  132825  &  44985  &  49176 & 14091  & 15239\\
 0.9 &   57185  &   64125  &  19694  &  23616 &  6205  &  7355\\
 1.0 &   -6323  &    -644  &  -4038  &   -487 & -1353  &  -176\\
 1.1 &  -63409  &  -59764  & -25481  & -22161 & -7988  & -6988\\
 1.2 & -113185  & -109888  & -43670  & -41073 &-13831  &-12873\\
 1.3 & -153819  & -151472  & -59174  & -56711 &-18649  &-17694\\
 1.4 & -186058  & -183899  & -70850  & -68593 &-22263  &-21514\\
 1.5 & -208250  & -207196  & -78835  & -77001 &-24713  &-23963\\
 1.6 & -221960  & -220977  & -83970  & -82376 &-26366  &-25744\\
 1.7 & -227736  & -226528  & -85950  & -84319 &-26923  &-26298\\
 1.8 & -226079  & -224735  & -85257  & -83785 &-26776  &-26311\\
 1.9 & -218582  & -217560  & -82259  & -81045 &-25838  &-25425\\
 2.0 & -206241  & -205534  & -77367  & -76353 &-24369  &-23956\\
 2.1 & -190950  & -190078  & -71645  & -70656 &-22425  &-22139\\
 2.2 & -173112  & -172168  & -64801  & -64083 &-20325  &-20044\\
 2.3 & -153979  & -152990  & -57654  & -57118 &-18180  &-18030\\
 2.4 & -134751  & -133959  & -50693  & -50169 &-16031  &-15874\\
 2.5 & -116288  & -115788  & -43807  & -43408 &-13815  &-13697\\
\hline 
\end{tabular}\end{center}
\tablecomments{$G_{\rm m,r}|_{z=32}$ and $G_{\rm m,r}|_{z=0}$ are
real-space genus values of the dark matter distributions at
initial epoch of the simulation ($z=32$) and at final epoch ($z=0$), respectively, 
using a large array of $2160^3$ pixels ($p=3.33\hMpc$).}
\end{table*}

\begin{table*}
\caption{
Genus Values at a Given Threshold Level for the Samples
Used to Estimate the Redshift-space Distortion Effects
}
\label{Tab:RSD}
\centering
\tiny
\begin{center}
\begin{tabular}{r|rr|rr|rr|rr|rr|rr}
\hline\hline
 &\multicolumn{6}{c|}{Matter} &\multicolumn{6}{c}{Halo} \\
 \cline{2-7}\cline{8-13}
 &\multicolumn{2}{c|}{$R_G=15h^{-1}$Mpc} &\multicolumn{2}{c|}{$R_G=22h^{-1}$Mpc} &\multicolumn{2}{c|}{$R_G=34h^{-1}$Mpc}
 &\multicolumn{2}{c|}{$R_G=15h^{-1}$Mpc} &\multicolumn{2}{c|}{$R_G=22h^{-1}$Mpc} &\multicolumn{2}{c}{$R_G=34h^{-1}$Mpc}\\
 \cline{2-3}\cline{4-5}\cline{6-7} \cline{8-9}\cline{10-11}\cline{12-13}
$\nu$ & $G_{\rm m,r}$ & $G_{\rm m,z}$
      & $G_{\rm m,r}$ & $G_{\rm m,z}$
      & $G_{\rm m,r}$ & $G_{\rm m,z}$ 
      & $G_{\rm h,r}$ & $G_{\rm h,z}$
      & $G_{\rm h,r}$ & $G_{\rm h,z}$
      & $G_{\rm h,r}$ & $G_{\rm h,z}$ \\
\hline
-2.5& -116313  & -105435  & -43267  & -39876 &-13683 &-12876 & -102261  & -99669  & -39240  & -38419  & -13003  & -12745   \\  
-2.4& -135266  & -122203  & -50342  & -46292 &-15901 &-14877 & -117627  & -114673  & -45125  & -44111  & -14855  & -14558  \\  
-2.3& -154571  & -139699  & -57606  & -52889 &-18078 &-16884 & -132824  & -129274  & -50919  & -49625  & -16752  & -16548  \\  
-2.2& -173653  & -156927  & -64555  & -59259 &-20234 &-18913 & -147195  & -143294  & -56411  & -55188  & -18574  & -18282  \\  
-2.1& -190606  & -173381  & -71101  & -65410 &-22285 &-20856 & -160452  & -156276  & -61145  & -59716  & -20149  & -19871  \\  
-2.0& -206042  & -187213  & -76917  & -70613 &-24115 &-22577 & -170947  & -166189  & -64882  & -63491  & -21432  & -21074  \\  
-1.9& -218159  & -197904  & -81628  & -74642 &-25448 &-23660 & -177276  & -172432  & -67059  & -65851  & -22160  & -21855  \\  
-1.8& -225541  & -204475  & -84236  & -77105 &-26240 &-24384 & -179216  & -174259  & -67728  & -66217  & -22333  & -21948  \\  
-1.7& -226813  & -205789  & -84723  & -77466 &-26353 &-24505 & -174754  & -170281  & -65966  & -64347  & -21698  & -21373  \\  
-1.6& -220701  & -200599  & -82486  & -75297 &-25706 &-23864 & -163768  & -159279  & -61641  & -60027  & -20253  & -19902  \\  
-1.5& -207088  & -188090  & -77269  & -70301 &-23980 &-22100 & -145480  & -141253  & -54512  & -53030  & -17983  & -17760  \\  
-1.4& -184193  & -167012  & -68525  & -62021 &-21381 &-19444 & -118359  & -115328  & -44231  & -43271  & -14680  & -14394  \\  
-1.3& -151186  & -137667  & -56390  & -50503 &-17438 &-15984 & -83337  & -80739  & -30665  & -29771  & -10304  & -10157    \\ 
-1.2& -109502  &  -99820  & -40876  & -35688 &-12575 &-11067 & -39293  & -38560  & -14141  & -13600  & -5102  & -4975      \\ 
-1.1&  -58746  &  -53195  & -21605  & -18061 & -6745 & -5385 & 10781  & 10990  & 5592  & 5403  & 1005  & 1191              \\ 
-1.0&    -307  &    -289  &   -104  &   2390 &   -68 &  1054 & 67923  & 66761  & 27243  & 26872  & 8260  & 8226            \\ 
-0.9&   64655  &   58683  &  23958  &  24762 &  7232 &  8123 & 129579  & 126265  & 50801  & 50347  & 15829  & 15854        \\ 
-0.8&  133567  &  121533  &  49094  &  48751 & 15138 & 15507 & 194007  & 189367  & 75305  & 73879  & 23590  & 23388        \\ 
-0.7&  203207  &  186681  &  75396  &  73429 & 23416 & 23154 & 258434  & 251997  & 99154  & 97575  & 31563  & 31110        \\ 
-0.6&  272365  &  248968  & 101707  &  97331 & 31371 & 30880 & 319970  & 311566  & 121886  & 120025  & 38994  & 38468      \\ 
-0.5&  337302  &  308522  & 125978  & 119804 & 39205 & 38090 & 375748  & 366234  & 142869  & 140126  & 45453  & 45045      \\ 
-0.4&  395286  &  361770  & 147247  & 139414 & 46214 & 44521 & 423008  & 412037  & 160525  & 157737  & 51101  & 50594      \\ 
-0.3&  444378  &  406444  & 164908  & 155752 & 51848 & 49977 & 459398  & 447844  & 174053  & 171113  & 55801  & 54976      \\ 
-0.2&  480671  &  439533  & 178298  & 168028 & 56050 & 53761 & 484631  & 472590  & 182765  & 179387  & 58512  & 57823      \\ 
-0.1&  503357  &  460455  & 186729  & 175348 & 58542 & 56116 & 495940  & 483765  & 186476  & 182929  & 59865  & 59131      \\ 
 0.0&  510839  &  467357  & 190051  & 177809 & 59482 & 56827 & 492597  & 479888  & 184577  & 181032  & 59107  & 58348      \\ 
 0.1&  503491  &  460428  & 186997  & 174608 & 58662 & 55971 & 473891  & 461746  & 177239  & 173748  & 56758  & 56031      \\ 
 0.2&  480720  &  439595  & 178788  & 166731 & 56074 & 53274 & 441327  & 430011  & 164478  & 161223  & 52722  & 52003      \\ 
 0.3&  444352  &  406060  & 165242  & 154089 & 51751 & 49126 & 394982  & 385197  & 146916  & 144258  & 46948  & 46315      \\ 
 0.4&  395592  &  361476  & 147385  & 136789 & 45881 & 43455 & 340326  & 331818  & 125546  & 123116  & 39850  & 39431      \\ 
 0.5&  337400  &  308017  & 125919  & 116257 & 38921 & 36614 & 277462  & 269690  & 101133  & 99298  & 32045  & 31555       \\ 
 0.6&  272285  &  247690  & 101880  &  93354 & 31760 & 29220 & 208490  & 202517  & 74849  & 73390  & 23701  & 23321        \\ 
 0.7&  203385  &  184505  &  75822  &  68442 & 23580 & 21473 & 136121  & 132933  & 47330  & 46327  & 14891  & 14660        \\ 
 0.8&  132825  &  119637  &  49176  &  43495 & 15239 & 13514 & 64902  & 63549  & 20326  & 19869  & 6462  & 6305            \\ 
 0.9&   64125  &   55910  &  23616  &  18879 &  7355 &  5810 & -3867  & -2808  & -5355  & -5415  & -1960  & -1874          \\ 
 1.0&    -644  &   -4157  &   -487  &  -4042 &  -176 & -1369 & -66799  & -65007  & -28645  & -28365  & -9337  & -9346      \\ 
 1.1&  -59764  &  -58193  & -22161  & -24271 & -6988 & -7838 & -122715  & -119034  & -49424  & -48931  & -15834  & -15720  \\  
 1.2& -109888  & -104589  & -41073  & -41814 &-12873 &-13565 & -170334  & -164877  & -67092  & -65886  & -21350  & -21214  \\  
 1.3& -151472  & -143835  & -56711  & -56487 &-17694 &-18231 & -207441  & -201443  & -80908  & -79406  & -25789  & -25545  \\  
 1.4& -183899  & -173656  & -68593  & -67831 &-21514 &-21736 & -235223  & -228670  & -90900  & -89482  & -28871  & -28728  \\  
 1.5& -207196  & -194941  & -77001  & -75753 &-23963 &-24046 & -252821  & -246087  & -97174  & -95306  & -31018  & -30740  \\  
 1.6& -220977  & -207743  & -82376  & -80344 &-25744 &-25614 & -261786  & -254261  & -100085  & -98463  & -31873 & -31574  \\ 
 1.7& -226528  & -212588  & -84319  & -82188 &-26298 &-26047 & -262781  & -255334  & -100080  & -98410  & -31775 & -31534  \\ 
 1.8& -224735  & -210939  & -83785  & -81382 &-26311 &-25910 & -256944  & -249623  & -97305  & -95670  & -31079  & -30688  \\  
 1.9& -217560  & -20000   & -81045  & -78645 &-25425 &-25009 & -245387  & -238300  & -92841  & -91152  & -29580  & -29273  \\  
 2.0& -205534  & -192481  & -76353  & -74198 &-23956 &-23591 & -228725  & -222284  & -86288  & -84666  & -27555  & -27282  \\  
 2.1& -190078  & -178068  & -70656  & -68258 &-22139 &-21790 & -209709  & -203700  & -78719  & -77300  & -25177  & -24899  \\  
 2.2& -172168  & -161285  & -64083  & -61821 &-20044 &-19774 & -188317  & -182962  & -70751  & -69411  & -22541  & -22399  \\  
 2.3& -152990  & -143461  & -57118  & -55137 &-18030 &-17615 & -166473  & -161865  & -62751  & -61617  & -19915  & -19663  \\  
 2.4& -133959  & -125628  & -50169  & -48402 &-15874 &-15474 & -144722  & -140654  & -54656  & -53737  & -17405  & -17220  \\  
 2.5& -115788  & -108441  & -43408  & -41751 &-13697 &-13353 & -124021  & -120785  & -46823  & -45995  & -14957  & -14800  \\  
\hline
\end{tabular}\end{center}
\tablecomments{$G_{\rm m,r}$ and $G_{\rm m,z}$ are
genus values of dark matter density fields 
in redshift space and real space at $z=0$, respectively.
$G_{\rm h,r}$ and $G_{\rm h,z}$ are
genus values of halo density fields in redshift space and real space at $z=0$, respectively.
A large array of $2160^3$ pixels ($p=3.33\hMpc$) is used.}
\end{table*}

\begin{deluxetable}{rrrrrrrrrr}
\tabletypesize{\tiny}
\tablecaption{
Genus Values at a Given Threshold Level for the Samples
to Estimate the Effects of Shot Noise
in the Matter Density Field \label{Tab:shotm}}
\tablewidth{0pt}
\tablehead{
\colhead{$\nu$} &\colhead{$\bar{d}/R_{G}=0.08$} & \colhead{0.16} & \colhead{0.32}&
\colhead{0.64}  &\colhead{1.0}     & \colhead{$\sqrt{2}$} & \colhead{$\sqrt{3}$} &
\colhead{2.0}   &\colhead{$\sqrt{5}$} 
}
\startdata
\cutinhead{$R_G=15\hMpc$}
-2.5 & -108803 & -109097 & -110747 & -120821 & -140353 & -154544 & -152069 & -140914 & -129580 \\
-2.4 & -126479 & -126635 & -128280 & -140073 & -161425 & -178082 & -175107 & -162247 & -148798 \\
-2.3 & -144533 & -144903 & -146578 & -158972 & -182971 & -202184 & -198173 & -183704 & -167654 \\
-2.2 & -162671 & -162865 & -164723 & -178013 & -203321 & -224608 & -220301 & -204104 & -186106 \\
-2.1 & -179447 & -179682 & -181609 & -195462 & -222438 & -245411 & -240344 & -222120 & -201790 \\
-2.0 & -193774 & -194052 & -196385 & -210250 & -238326 & -261728 & -256636 & -235974 & -213593 \\
-1.9 & -205192 & -205588 & -207647 & -221167 & -249510 & -272704 & -267465 & -245113 & -220938 \\
-1.8 & -212516 & -212808 & -214623 & -227601 & -254942 & -276779 & -270455 & -246928 & -220084 \\
-1.7 & -214418 & -214488 & -215837 & -228118 & -253665 & -271836 & -264220 & -239035 & -211509 \\
-1.6 & -208895 & -208873 & -210525 & -220441 & -242145 & -256844 & -246865 & -220657 & -192357 \\
-1.5 & -195759 & -195714 & -197125 & -204527 & -220545 & -230384 & -216905 & -190261 & -161745 \\
-1.4 & -174307 & -174532 & -175099 & -179766 & -189677 & -191601 & -174263 & -146521 & -118847 \\
-1.3 & -143465 & -143340 & -143716 & -145065 & -146893 & -140741 & -119294 & -90084 & -63894 \\
-1.2 & -103662 & -103518 & -103184 & -101097 & -94008 & -76548 & -50740 & -20089 & 4722 \\
-1.1 & -56031 & -55991 & -54581 & -47947 & -31315 & -2501 & 28810 & 61627 & 84663 \\
-1.0 & -503 & -519 & 1661 & 12767 & 40229 & 81058 & 118224 & 153571 & 174303 \\
-0.9 & 60775 & 61199 & 63872 & 79433 & 118138 & 171146 & 214279 & 252995 & 272179 \\
-0.8 & 126761 & 127474 & 130070 & 149998 & 198890 & 264895 & 313851 & 358777 & 375211 \\
-0.7 & 194422 & 194879 & 197663 & 221578 & 279988 & 358710 & 413485 & 460638 & 477452 \\
-0.6 & 260443 & 261029 & 264630 & 292040 & 358638 & 449167 & 508048 & 557504 & 575569 \\
-0.5 & 322939 & 323275 & 327086 & 358595 & 430979 & 531361 & 594576 & 644719 & 664937 \\
-0.4 & 378167 & 378349 & 383541 & 416462 & 493846 & 601666 & 669233 & 718003 & 736801 \\
-0.3 & 425068 & 425911 & 430474 & 462874 & 544638 & 655540 & 726074 & 772234 & 792788 \\
-0.2 & 460248 & 461023 & 465314 & 498348 & 580280 & 692958 & 762495 & 806075 & 833702 \\
-0.1 & 482725 & 483160 & 486930 & 519289 & 598023 & 709078 & 776693 & 817687 & 851076 \\
0.0 & 490162 & 490602 & 494963 & 524479 & 599075 & 704346 & 768197 & 805065 & 839598 \\
0.1 & 482659 & 483270 & 486690 & 513908 & 582957 & 677895 & 737392 & 769167 & 801019 \\
0.2 & 461251 & 461386 & 465121 & 488200 & 548823 & 633128 & 683867 & 711796 & 738378 \\
0.3 & 426363 & 426333 & 429141 & 448002 & 499130 & 569389 & 611928 & 633861 & 652975 \\
0.4 & 379233 & 379263 & 381873 & 396700 & 436841 & 491824 & 524140 & 540298 & 552158 \\
0.5 & 322841 & 323181 & 324791 & 335245 & 363431 & 403392 & 425264 & 435647 & 437450 \\
0.6 & 260225 & 260348 & 261250 & 267166 & 282856 & 307069 & 319402 & 322174 & 315898 \\
0.7 & 192897 & 192901 & 193653 & 195594 & 200500 & 207587 & 209686 & 204974 & 192795 \\
0.8 & 124318 & 124534 & 124193 & 122231 & 116915 & 110098 & 101429 & 90069 & 72889 \\
0.9 & 56915 & 56629 & 55862 & 51428 & 37036 & 16423 & -2675 & -19102 & -39319 \\
1.0 & -6521 & -6545 & -8008 & -15212 & -37125 & -70316 & -96892 & -118535 & -139294 \\
1.1 & -63131 & -63701 & -64788 & -75217 & -102750 & -145860 & -180410 & -205552 & -226705 \\
1.2 & -113080 & -113102 & -114836 & -126726 & -158058 & -209987 & -249572 & -278323 & -301021 \\
1.3 & -153644 & -154289 & -156083 & -169290 & -203187 & -260274 & -303650 & -335258 & -359262 \\
1.4 & -185906 & -185649 & -187593 & -201015 & -237710 & -297518 & -342879 & -376007 & -400979 \\
1.5 & -208150 & -208352 & -209985 & -223703 & -260458 & -321559 & -368248 & -401699 & -426925 \\
1.6 & -221915 & -222178 & -223476 & -236510 & -272425 & -333810 & -379409 & -412845 & -438066 \\
1.7 & -227557 & -227578 & -229451 & -241591 & -275986 & -334481 & -379473 & -411791 & -435412 \\
1.8 & -225816 & -225698 & -227421 & -239490 & -271383 & -326114 & -368835 & -399876 & -422136 \\
1.9 & -218403 & -218343 & -219950 & -230627 & -259817 & -310545 & -350630 & -378792 & -399210 \\
2.0 & -205990 & -206152 & -207552 & -217543 & -243245 & -289090 & -326096 & -352110 & -370887 \\
2.1 & -190671 & -190841 & -191736 & -200667 & -223458 & -264538 & -297114 & -321526 & -337949 \\
2.2 & -172966 & -173058 & -173890 & -181509 & -201590 & -237259 & -265725 & -287890 & -301764 \\
2.3 & -153895 & -154013 & -154748 & -161439 & -178603 & -209352 & -233754 & -252854 & -265216 \\
2.4 & -134555 & -134694 & -135611 & -141081 & -155689 & -181824 & -202628 & -218861 & -229496 \\
2.5 & -116095 & -116228 & -116613 & -121666 & -133372 & -155784 & -172944 & -186684 & -195217 \\
\cutinhead{$R_G=22\hMpc$}
-2.5 & -40688 & -109097 & -110747 & -120821 & -140353 & -154544 & -152069 & -140914 & -129580 \\
-2.4 & -47089 & -126635 & -128280 & -140073 & -161425 & -178082 & -175107 & -162247 & -148798 \\
-2.3 & -53783 & -144903 & -146578 & -158972 & -182971 & -202184 & -198173 & -183704 & -167654 \\
-2.2 & -60464 & -162865 & -164723 & -178013 & -203321 & -224608 & -220301 & -204104 & -186106 \\
-2.1 & -66691 & -179682 & -181609 & -195462 & -222438 & -245411 & -240344 & -222120 & -201790 \\
-2.0 & -72159 & -194052 & -196385 & -210250 & -238326 & -261728 & -256636 & -235974 & -213593 \\
-1.9 & -76305 & -205588 & -207647 & -221167 & -249510 & -272704 & -267465 & -245113 & -220938 \\
-1.8 & -78548 & -212808 & -214623 & -227601 & -254942 & -276779 & -270455 & -246928 & -220084 \\
-1.7 & -78965 & -214488 & -215837 & -228118 & -253665 & -271836 & -264220 & -239035 & -211509 \\
-1.6 & -76721 & -208873 & -210525 & -220441 & -242145 & -256844 & -246865 & -220657 & -192357 \\
-1.5 & -71404 & -195714 & -197125 & -204527 & -220545 & -230384 & -216905 & -190261 & -161745 \\
-1.4 & -63360 & -174532 & -175099 & -179766 & -189677 & -191601 & -174263 & -146521 & -118847 \\
-1.3 & -51538 & -143340 & -143716 & -145065 & -146893 & -140741 & -119294 & -90084 & -63894 \\
-1.2 & -36399 & -103518 & -103184 & -101097 & -94008 & -76548 & -50740 & -20089 & 4722 \\
-1.1 & -17998 & -55991 & -54581 & -47947 & -31315 & -2501 & 28810 & 61627 & 84663 \\
-1.0 & 2571 & -519 & 1661 & 12767 & 40229 & 81058 & 118224 & 153571 & 174303 \\
-0.9 & 25090 & 61199 & 63872 & 79433 & 118138 & 171146 & 214279 & 252995 & 272179 \\
-0.8 & 49339 & 127474 & 130070 & 149998 & 198890 & 264895 & 313851 & 358777 & 375211 \\
-0.7 & 73866 & 194879 & 197663 & 221578 & 279988 & 358710 & 413485 & 460638 & 477452 \\
-0.6 & 98822 & 261029 & 264630 & 292040 & 358638 & 449167 & 508048 & 557504 & 575569 \\
-0.5 & 121655 & 323275 & 327086 & 358595 & 430979 & 531361 & 594576 & 644719 & 664937 \\
-0.4 & 142333 & 378349 & 383541 & 416462 & 493846 & 601666 & 669233 & 718003 & 736801 \\
-0.3 & 159218 & 425911 & 430474 & 462874 & 544638 & 655540 & 726074 & 772234 & 792788 \\
-0.2 & 171894 & 461023 & 465314 & 498348 & 580280 & 692958 & 762495 & 806075 & 833702 \\
-0.1 & 179482 & 483160 & 486930 & 519289 & 598023 & 709078 & 776693 & 817687 & 851076 \\
0.0 & 181657 & 490602 & 494963 & 524479 & 599075 & 704346 & 768197 & 805065 & 839598 \\
0.1 & 178930 & 483270 & 486690 & 513908 & 582957 & 677895 & 737392 & 769167 & 801019 \\
0.2 & 170814 & 461386 & 465121 & 488200 & 548823 & 633128 & 683867 & 711796 & 738378 \\
0.3 & 157726 & 426333 & 429141 & 448002 & 499130 & 569389 & 611928 & 633861 & 652975 \\
0.4 & 140133 & 379263 & 381873 & 396700 & 436841 & 491824 & 524140 & 540298 & 552158 \\
0.5 & 119100 & 323181 & 324791 & 335245 & 363431 & 403392 & 425264 & 435647 & 437450 \\
0.6 & 95226 & 260348 & 261250 & 267166 & 282856 & 307069 & 319402 & 322174 & 315898 \\
0.7 & 69910 & 192901 & 193653 & 195594 & 200500 & 207587 & 209686 & 204974 & 192795 \\
0.8 & 44396 & 124534 & 124193 & 122231 & 116915 & 110098 & 101429 & 90069 & 72889 \\
0.9 & 19698 & 56629 & 55862 & 51428 & 37036 & 16423 & -2675 & -19102 & -39319 \\
1.0 & -4087 & -6545 & -8008 & -15212 & -37125 & -70316 & -96892 & -118535 & -139294 \\
1.1 & -25061 & -63701 & -64788 & -75217 & -102750 & -145860 & -180410 & -205552 & -226705 \\
1.2 & -43207 & -113102 & -114836 & -126726 & -158058 & -209987 & -249572 & -278323 & -301021 \\
1.3 & -58281 & -154289 & -156083 & -169290 & -203187 & -260274 & -303650 & -335258 & -359262 \\
1.4 & -69736 & -185649 & -187593 & -201015 & -237710 & -297518 & -342879 & -376007 & -400979 \\
1.5 & -77986 & -208352 & -209985 & -223703 & -260458 & -321559 & -368248 & -401699 & -426925 \\
1.6 & -82798 & -222178 & -223476 & -236510 & -272425 & -333810 & -379409 & -412845 & -438066 \\
1.7 & -84849 & -227578 & -229451 & -241591 & -275986 & -334481 & -379473 & -411791 & -435412 \\
1.8 & -84219 & -225698 & -227421 & -239490 & -271383 & -326114 & -368835 & -399876 & -422136 \\
1.9 & -81147 & -218343 & -219950 & -230627 & -259817 & -310545 & -350630 & -378792 & -399210 \\
2.0 & -76466 & -206152 & -207552 & -217543 & -243245 & -289090 & -326096 & -352110 & -370887 \\
2.1 & -70787 & -190841 & -191736 & -200667 & -223458 & -264538 & -297114 & -321526 & -337949 \\
2.2 & -64074 & -173058 & -173890 & -181509 & -201590 & -237259 & -265725 & -287890 & -301764 \\
2.3 & -56978 & -154013 & -154748 & -161439 & -178603 & -209352 & -233754 & -252854 & -265216 \\
2.4 & -50072 & -134694 & -135611 & -141081 & -155689 & -181824 & -202628 & -218861 & -229496 \\
2.5 & -43497 & -116228 & -116613 & -121666 & -133372 & -155784 & -172944 & -186684 & -195217 \\
\cutinhead{$R_G=22\hMpc$}
-2.5 & -12844 & -13354 & -13600 & -15337 & -16780 & -16807 & -15401 & -14119 & -12612 \\
-2.4 & -14854 & -15431 & -15829 & -17695 & -19430 & -19342 & -17876 & -16288 & -14565 \\
-2.3 & -16849 & -17484 & -17942 & -20206 & -22028 & -21906 & -20329 & -18395 & -16472 \\
-2.2 & -18856 & -19555 & -20054 & -22606 & -24555 & -24530 & -22562 & -20374 & -18214 \\
-2.1 & -20729 & -21610 & -22013 & -24734 & -26864 & -26714 & -24672 & -22257 & -19815 \\
-2.0 & -22274 & -23239 & -23658 & -26660 & -28919 & -28559 & -26452 & -23712 & -21042 \\
-1.9 & -23563 & -24606 & -25002 & -27868 & -30160 & -29912 & -27449 & -24744 & -21776 \\
-1.8 & -24383 & -25300 & -25774 & -28504 & -30899 & -30523 & -27934 & -24806 & -21744 \\
-1.7 & -24319 & -25472 & -25911 & -28327 & -30678 & -30175 & -27275 & -24114 & -20875 \\
-1.6 & -23552 & -24593 & -25113 & -27332 & -29345 & -28587 & -25878 & -22394 & -19078 \\
-1.5 & -22029 & -22755 & -23265 & -25192 & -26906 & -26048 & -23201 & -19523 & -16178 \\
-1.4 & -19384 & -20257 & -20447 & -21838 & -23265 & -22006 & -18939 & -15378 & -12329 \\
-1.3 & -15861 & -16287 & -16520 & -17431 & -18081 & -16472 & -13409 & -9740 & -7148 \\
-1.2 & -11143 & -11287 & -11737 & -11725 & -11464 & -9641 & -6630 & -2985 & -480 \\
-1.1 & -5672 & -5504 & -5528 & -4882 & -3675 & -1452 & 1517 & 5031 & 7150 \\
-1.0 & 798 & 975 & 1243 & 2797 & 4987 & 7608 & 10663 & 14062 & 15805 \\
-0.9 & 7885 & 8342 & 8691 & 10976 & 14479 & 17511 & 20559 & 23967 & 25243 \\
-0.8 & 15496 & 16038 & 16749 & 19943 & 24403 & 27965 & 30638 & 33980 & 34978 \\
-0.7 & 22911 & 23809 & 24935 & 29156 & 34474 & 38313 & 40732 & 44244 & 44712 \\
-0.6 & 30511 & 31933 & 33051 & 38183 & 44454 & 48555 & 50774 & 53895 & 54360 \\
-0.5 & 37709 & 39351 & 40376 & 46529 & 53505 & 57923 & 60067 & 63004 & 63186 \\
-0.4 & 44138 & 45966 & 47240 & 53677 & 61224 & 65991 & 68138 & 70686 & 70306 \\
-0.3 & 49717 & 51460 & 52750 & 59485 & 67650 & 72718 & 74317 & 76483 & 75982 \\
-0.2 & 53569 & 55728 & 56774 & 63845 & 72250 & 77222 & 78558 & 79944 & 80001 \\
-0.1 & 56079 & 58040 & 59463 & 66086 & 74762 & 79241 & 80715 & 81341 & 82071 \\
0.0 & 56748 & 58696 & 60024 & 66780 & 75083 & 79270 & 80382 & 80680 & 81265 \\
0.1 & 55917 & 57799 & 58750 & 65380 & 72879 & 77103 & 77705 & 77567 & 78072 \\
0.2 & 53226 & 55093 & 55930 & 61973 & 68813 & 72300 & 72770 & 71943 & 72383 \\
0.3 & 49125 & 50490 & 51493 & 56736 & 63083 & 65625 & 65741 & 64824 & 64486 \\
0.4 & 43441 & 44502 & 45616 & 49795 & 55189 & 56828 & 56840 & 55703 & 54977 \\
0.5 & 36605 & 37709 & 38545 & 42081 & 45846 & 47180 & 46378 & 45179 & 44267 \\
0.6 & 29505 & 30192 & 30702 & 33329 & 35881 & 36448 & 35171 & 33950 & 32573 \\
0.7 & 21717 & 22272 & 22466 & 24126 & 25489 & 25147 & 23655 & 22115 & 20591 \\
0.8 & 13793 & 14235 & 14180 & 14782 & 15164 & 13895 & 12239 & 10780 & 8782 \\
0.9 & 5987 & 6218 & 6056 & 5667 & 5072 & 3100 & 1230 & -318 & -2268 \\
1.0 & -1291 & -1401 & -1722 & -2848 & -4273 & -7017 & -8777 & -10499 & -12437 \\
1.1 & -7738 & -8162 & -8390 & -10444 & -12805 & -16029 & -17850 & -19454 & -21313 \\
1.2 & -13389 & -13960 & -14285 & -16936 & -20202 & -23410 & -25605 & -27246 & -28571 \\
1.3 & -18018 & -18639 & -19267 & -22124 & -26016 & -29747 & -31646 & -33377 & -34223 \\
1.4 & -21607 & -22290 & -23003 & -26024 & -30662 & -34246 & -36180 & -37769 & -38628 \\
1.5 & -24030 & -24923 & -25469 & -28788 & -33533 & -37203 & -39110 & -40538 & -41639 \\
1.6 & -25565 & -26425 & -26931 & -30303 & -35316 & -38944 & -40786 & -42049 & -42804 \\
1.7 & -26322 & -26969 & -27596 & -30870 & -35865 & -39253 & -41329 & -42400 & -42921 \\
1.8 & -26092 & -26771 & -27300 & -30504 & -35319 & -38599 & -40292 & -41483 & -42070 \\
1.9 & -25182 & -25903 & -26386 & -29434 & -34071 & -37024 & -38520 & -39493 & -39948 \\
2.0 & -23804 & -24471 & -24849 & -27742 & -31838 & -34712 & -36128 & -36930 & -37405 \\
2.1 & -22021 & -22450 & -23008 & -25534 & -29360 & -31972 & -33090 & -33809 & -34229 \\
2.2 & -19907 & -20408 & -20922 & -23249 & -26563 & -28807 & -29757 & -30457 & -30751 \\
2.3 & -17707 & -18202 & -18678 & -20717 & -23673 & -25566 & -26288 & -26911 & -27347 \\
2.4 & -15613 & -16077 & -16455 & -18181 & -20679 & -22445 & -22866 & -23527 & -23790 \\
2.5 & -13536 & -13970 & -14218 & -15665 & -17856 & -19326 & -19668 & -20186 & -20380 \\
\enddata
\tablecomments{Genus values at a given threshold level of the density field estimated from randomly sampled
simulation particles to have mean particle separations of
$\bar{d}/R=$0.08, 0.16, 0.32, 0.64, 1.0, $\sqrt{2}$, $\sqrt{3}$, 2.0, $\sqrt{5}$
using the smallest possible pixel size ($2160^3$ in this case).}
\end{deluxetable}

\begin{deluxetable}{rrrrrr}
\tablewidth{0pt}
\tablecaption{
Genus Values at a Given Threshold Level for the Samples
to Estimate the Effects of Shot Noise
in the Halo Density Field \label{Tab:shoth}}
\tabletypesize{\tiny}
\tablehead{
\colhead{$\nu$} &\colhead{$\bar{d}/R_{G}=1.0$} &
\colhead{$\sqrt{2}$} & \colhead{$\sqrt{3}$} &
\colhead{2.0}   &\colhead{$\sqrt{5}$} 
}
\startdata
\cutinhead{$R_G=15\hMpc$}
-2.5 & -102261 & -108020 & -109474 & -106170 & -101833\\
-2.4 & -117627 & -123827 & -125504 & -121650 & -116447\\
-2.3 & -132824 & -139315 & -140723 & -137159 & -130756\\
-2.2 & -147195 & -154361 & -155424 & -151801 & -144163\\
-2.1 & -160452 & -167495 & -168485 & -163702 & -155013\\
-2.0 & -170947 & -177252 & -177619 & -172557 & -162893\\
-1.9 & -177276 & -182981 & -182718 & -176658 & -166152\\
-1.8 & -179216 & -182780 & -182135 & -175039 & -163838\\
-1.7 & -174754 & -176636 & -174517 & -166667 & -154063\\
-1.6 & -163768 & -162520 & -158629 & -149564 & -136685\\
-1.5 & -145480 & -140541 & -133559 & -123369 & -110014\\
-1.4 & -118359 & -109849 & -100048 &  -87422 &  -73636\\
-1.3 &  -83337 &  -70361 &  -56574 &  -42703 &  -26844\\
-1.2 &  -39293 &  -22161 &   -4746 &   12335 &   29660\\
-1.1 &   10781 &   33457 &   55363 &   75117 &   94038\\
-1.0 &   67923 &   95597 &  121050 &  143155 &  162763\\
-0.9 &  129579 &  161808 &  190883 &  215662 &  236596\\
-0.8 &  194007 &  230086 &  262897 &  289390 &  310610\\
-0.7 &  258434 &  297029 &  331883 &  361392 &  382265\\
-0.6 &  319970 &  360435 &  397567 &  427809 &  447209\\
-0.5 &  375748 &  418230 &  455066 &  484099 &  504355\\
-0.4 &  423008 &  465165 &  501746 &  530790 &  549739\\
-0.3 &  459398 &  501054 &  534786 &  562431 &  579517\\
-0.2 &  484631 &  522542 &  553703 &  578904 &  592422\\
-0.1 &  495940 &  528805 &  555779 &  576657 &  588063\\
 0.0 &  492597 &  518380 &  540967 &  556871 &  564573\\
 0.1 &  473891 &  492472 &  509659 &  520117 &  523738\\
 0.2 &  441327 &  451759 &  462218 &  466762 &  466319\\
 0.3 &  394982 &  398062 &  402301 &  399666 &  395809\\
 0.4 &  340326 &  333713 &  329257 &  322895 &  315912\\
 0.5 &  277462 &  260998 &  248828 &  237276 &  227391\\
 0.6 &  208490 &  183080 &  164819 &  148803 &  135606\\
 0.7 &  136121 &  103644 &   79499 &   59913 &   44572\\
 0.8 &   64902 &   25544 &   -3520 &  -26164 &  -42887\\
 0.9 &   -3867 &  -47742 &  -80447 & -105376 & -123637\\
 1.0 &  -66799 & -115078 & -150026 & -175245 & -194174\\
 1.1 &  -122715 & -173527 & -209089 & -234217 & -253666\\
 1.2 &  -170334 & -221263 & -257194 & -282181 & -299266\\
 1.3 & -207441 & -258785 & -293146 & -316599 & -331726\\
 1.4 & -235223 & -285054 & -317405 & -338436 & -351558\\
 1.5 & -252821 & -300189 & -329796 & -348415 & -359032\\
 1.6 & -261786 & -305509 & -331616 & -347459 & -355659\\
 1.7 & -262781 & -302178 & -324510 & -336669 & -343502\\
 1.8 & -256944 & -291340 & -309915 & -319156 & -323581\\
 1.9 & -245387 & -275040 & -289673 & -296387 & -298510\\
 2.0 & -228725 & -254128 & -265647 & -269897 & -269677\\
 2.1 & -209709 & -230806 & -238976 & -241136 & -239522\\
 2.2 & -188317 & -205504 & -211144 & -211829 & -209157\\
 2.3 & -166473 & -180325 & -183664 & -182938 & -179760\\
 2.4 & -144722 & -155422 & -156655 & -155250 & -151999\\
 2.5 & -124021 & -132169 & -131975 & -129767 & -126670\\
\cutinhead{$R_G=22\hMpc$}
-2.5 &-39240 & -41259 & -41109 & -39327 & -36984 \\
-2.4 &-45125 & -47043 & -47054 & -45075 & -42217 \\
-2.3 &-50919 & -53069 & -52780 & -50953 & -47455 \\
-2.2 &-56411 & -58272 & -58466 & -55967 & -52223 \\
-2.1 &-61145 & -63374 & -63524 & -60588 & -56409 \\
-2.0 &-64882 & -67341 & -67200 & -64109 & -59086 \\
-1.9 &-67059 & -69271 & -69026 & -65674 & -60484 \\
-1.8 &-67728 & -69454 & -68577 & -64794 & -59760 \\
-1.7 &-65966 & -67060 & -65555 & -61641 & -56524 \\
-1.6 &-61641 & -62013 & -60027 & -55782 & -50435 \\
-1.5 &-54512 & -53243 & -50637 & -46547 & -40904 \\
-1.4 &-44231 & -41785 & -37999 & -33344 & -27912 \\
-1.3 &-30665 & -26945 & -22160 & -16933 & -11368 \\
-1.2 &-14141 &  -8767 & -2800 & 3178 & 8265 \\
-1.1 &  5592 &  12530 & 19153 & 26188 & 31360 \\
-1.0 & 27243 &  36448 & 43587 & 51207 & 56317 \\
-0.9 & 50801 &  61563 & 69836 & 78177 & 83189 \\
-0.8 & 75305 &  87308 & 96906 & 105178 & 110616 \\
-0.7 & 99154 & 112420 & 122738 & 131487 & 136867 \\
-0.6 &121886 & 136108 & 147130 & 155495 & 160938 \\
-0.5 &142869 & 157547 & 168344 & 176823 & 181592 \\
-0.4 &160525 & 174935 & 185602 & 192945 & 198259 \\
-0.3 &174053 & 188207 & 197883 & 204058 & 209079 \\
-0.2 &182765 & 196347 & 204596 & 210383 & 214219 \\
-0.1 &186476 & 197740 & 206139 & 210769 & 213430 \\
 0.0 &184577 & 193817 & 200893 & 203528 & 205155 \\
 0.1 &177239 & 183928 & 189136 & 190869 & 190938 \\
 0.2 &164478 & 168156 & 171139 & 171460 & 170946 \\
 0.3 &146916 & 147916 & 148645 & 147412 & 146002 \\
 0.4 &125546 & 123652 & 122374 & 119459 & 116617 \\
 0.5 &101133 &  96432 & 93219 & 88841 & 84891 \\
 0.6 & 74849 &  67364 & 62297 & 56920 & 51770 \\
 0.7 & 47330 &  37850 & 30330 & 24670 & 18811 \\
 0.8 & 20326 &   9308 & -124 & -6147 & -12533 \\
 0.9 & -5355 & -18105 & -28083 & -34906 & -41649 \\
 1.0 &-28645 & -42811 & -53679 & -60894 & -67294 \\
 1.1 & -49424 & -64795 & -75767 & -83063 & -88872 \\
 1.2 & -67092 & -82611 & -93669 & -100550 & -106199 \\
 1.3 &-80908 & -96550 & -107076 & -113841 & -118475 \\
 1.4 &-90900 &-106081 & -116170 & -122405 & -126265 \\
 1.5 &-97174 &-111921 & -121026 & -126528 & -129264 \\
 1.6 &100085 &-113943 & -122360 & -126783 & -128716 \\
 1.7 &100080 &-113051 & -120183 & -123460 & -124694 \\
 1.8 &-97305 &-108877 & -115058 & -117521 & -118152 \\
 1.9 &-92841 &-102527 & -107607 & -109793 & -109814 \\
 2.0 &-86288 & -94916 & -98579 & -100293 & -99993 \\
 2.1 &-78719 & -86262 & -88788 & -89855 & -89278 \\
 2.2 &-70751 & -76817 & -78635 & -79231 & -78175 \\
 2.3 &-62751 & -67658 & -68476 & -68528 & -67542 \\
 2.4 &-54656 & -58547 & -58758 & -58571 & -57427 \\
 2.5 &-46823 & -49771 & -49782 & -49169 & -48194 \\
\cutinhead{$R_G=34\hMpc$}
-2.5 & -13003 & -13331 & -13040 & -12410 & -11284 \\
-2.4 & -14855 & -15434 & -14949 & -14116 & -12996 \\
-2.3 & -16752 & -17358 & -16779 & -15922 & -14629 \\
-2.2 & -18574 & -19165 & -18415 & -17569 & -16130 \\
-2.1 & -20149 & -20684 & -20014 & -19069 & -17350 \\
-2.0 & -21432 & -21861 & -21380 & -20081 & -18322 \\
-1.9 & -22160 & -22729 & -22002 & -20662 & -18817 \\
-1.8 & -22333 & -22850 & -22062 & -20632 & -18437 \\
-1.7 & -21698 & -22082 & -21461 & -19603 & -17600 \\
-1.6 & -20253 & -20407 & -19827 & -17937 & -15661 \\
-1.5 & -17983 & -17881 & -17046 & -15132 & -12864 \\
-1.4 & -14680 & -14215 & -13249 & -11211 & -9000 \\
-1.3 & -10304 & -9514 & -8122 & -5975 & -4365 \\
-1.2 & -5102 & -3738 & -2146 & 142 & 1631 \\
-1.1 & 1005 & 2955 & 4921 & 7094 & 8514 \\
-1.0 & 8260 & 10405 & 12707 & 14570 & 15996 \\
-0.9 & 15829 & 18394 & 20947 & 22775 & 23935 \\
-0.8 & 23590 & 26694 & 29452 & 31074 & 32053 \\
-0.7 & 31563 & 34943 & 37767 & 39350 & 40216 \\
-0.6 & 38994 & 42745 & 45582 & 46744 & 47957 \\
-0.5 & 45453 & 49689 & 52267 & 53517 & 54663 \\
-0.4 & 51101 & 55463 & 58003 & 58984 & 59819 \\
-0.3 & 55801 & 59482 & 62087 & 62933 & 63356 \\
-0.2 & 58512 & 62199 & 64391 & 65077 & 65149 \\
-0.1 & 59865 & 62805 & 64698 & 65070 & 65130 \\
0.0 & 59107 & 61839 & 63414 & 63010 & 62845 \\
0.1 & 56758 & 58983 & 60065 & 59689 & 59187 \\
0.2 & 52722 & 54259 & 54662 & 54115 & 53504 \\
0.3 & 46948 & 48247 & 47700 & 47126 & 45825 \\
0.4 & 39850 & 40422 & 39422 & 38493 & 37268 \\
0.5 & 32045 & 31724 & 30358 & 29220 & 27764 \\
0.6 & 23701 & 22500 & 20633 & 19506 & 17914 \\
0.7 & 14891 & 13017 & 10983 & 9542 & 7968 \\
0.8 & 6462 & 3769 & 1554 & 145 & -1783 \\
0.9 & -1960 & -4709 & -7158 & -9108 & -11023 \\
1.0 & -9337 & -12696 & -15386 & -16972 & -18989 \\
1.1 & -15834 & -19590 & -22466 & -24032 & -25592 \\
1.2 & -21350 & -25204 & -28120 & -29663 & -31254 \\
1.3 & -25789 & -29683 & -32332 & -33969 & -35262 \\
1.4 & -28871 & -32838 & -35265 & -36716 & -37851 \\
1.5 & -31018 & -34746 & -37004 & -38158 & -39063 \\
1.6 & -31873 & -35456 & -37565 & -38514 & -39273 \\
1.7 & -31775 & -35277 & -37080 & -37839 & -38406 \\
1.8 & -31079 & -33946 & -35419 & -36344 & -36496 \\
1.9 & -29580 & -32029 & -33483 & -34103 & -34085 \\
2.0 & -27555 & -29677 & -30998 & -31259 & -31187 \\
2.1 & -25177 & -26956 & -28029 & -28243 & -28017 \\
2.2 & -22541 & -24151 & -24925 & -24974 & -24821 \\
2.3 & -19915 & -21277 & -21783 & -21754 & -21564 \\
2.4 & -17405 & -18449 & -18679 & -18714 & -18456 \\
2.5 & -14957 & -15679 & -15821 & -15895 & -15487 \\
\enddata
\tablecomments{
Genus values at a given threshold level of the halo density field estimated from randomly selected
halos to have mean halo separations of
$\bar{d}/R=1.0$, $\sqrt{2}$, $\sqrt{3}$, 2.0, $\sqrt{5}$
using the smallest possible pixel size ($2160^3$ in this case).
}
\end{deluxetable}

\end{document}